\documentclass[11pt,a4paper]{article}
\usepackage{jcappub}

\usepackage{appendix}
\usepackage{array}
\newcolumntype{x}[1]{>{\centering\arraybackslash}p{#1}}

\def\lsim{\mathrel{\hbox{\rlap{\hbox{\lower4pt\hbox{$\sim$}}}\hbox{$<$}}}}

\makeatletter

\newcommand{\Rmnum}[1]{\expandafter\@slowromancap\romannumeral #1@}
\makeatother

\title{Aberration features in directional dark matter detection}

\author[a]{Nassim Bozorgnia,}
\author[a]{Graciela B. Gelmini}
\author[b,c]{and Paolo Gondolo}
\affiliation[a]{Department of Physics and Astronomy, UCLA,\\
475 Portola Plaza, Los Angeles, CA 90095, USA}
\affiliation[b]{Department of Physics and Astronomy, University of Utah,\\
115 South 1400 East \#201, Salt Lake City, UT 84112, USA}
\affiliation[c]{Department of Physics and Astronomy, Seoul National University,\\
Seoul, Korea 151-747}
\emailAdd{nassim@physics.ucla.edu}
\emailAdd{gelmini@physics.ucla.edu}
\emailAdd{paolo@physics.utah.edu}

\abstract{
The motion of the Earth around the Sun causes an annual change in the magnitude and direction of the arrival velocity of dark matter particles on Earth, in a way analogous to aberration of stellar light. In directional detectors, aberration of weakly interacting massive particles (WIMPs) modulates the pattern of nuclear recoil directions in a way that depends on the orbital velocity of the Earth and the local galactic distribution of WIMP velocities. Knowing the former, WIMP aberration can give information on the latter, besides being a curious way of confirming the revolution of the Earth and the extraterrestrial provenance of WIMPs. While observing the full aberration pattern requires extremely large exposures, we claim that the annual variation of the mean recoil direction or of the event counts over specific solid angles may be detectable with moderately large exposures. For example, integrated counts over Galactic hemispheres separated by planes perpendicular to Earth's orbit would modulate annually, resulting in Galactic Hemisphere Annual Modulations (GHAM) with amplitudes larger than the usual non-directional annual modulation.
}

\arxivnumber{1205.2333}

\begin{document}
\maketitle

\section{Introduction}

Direct dark matter experiments search for energy deposited by the scattering of Weakly Interacting Massive Particles (WIMPs) in the dark halo of our galaxy. In non-directional direct detection the only dark matter signature is  the annual modulation of the interaction rate due to the motion of the Earth around the Sun~\cite{Drukier}. This motion periodically changes the velocity  ${\bf V}_{\rm lab}$ of the detector with respect to the Galaxy, or equivalently the average velocity of the dark matter particles  with respect to the detector. If particles in the dark halo of our galaxy are on average at rest with respect to the Galaxy, the average velocity of WIMPs with respect to the detector is  just  $-{\bf V}_{\rm lab}$. This is the case of the Standard Halo Model. As in this model, we adopt an isotropic Maxwell-Boltzmann (IMB) velocity distribution, but with a dispersion and average velocity chosen to reflect our best knowledge of the dark halo of our galaxy.

In directional detectors~\cite{Ahlen:2009ev}, the two dark matter signatures easiest to detect are first a departure from isotropy  of the recoil directions with respect to the Galaxy, and second an average recoil direction in the direction of $-{\bf V}_{\rm lab}$~\cite{Spergel, Copi:1999pw, Alenazi-Gondolo:2008, Green&Morgan:2010, Morgan:2005,Green:2010gw}. Directional detectors, such as DRIFT~\cite{DRIFT}, DMTPC~\cite{DMTPC}, NEWAGE~\cite{NEWAGE} and MIMAC~\cite{MIMAC} use  CS$_2$, CF$_4$, $^3$He or other compounds. Refs.~\cite{Green:2010gw, Morgan:2005} find  that  for an S detector 9 events above 20 keV would be enough to reject isotropy, while between 27 and 32 events, depending on the velocity distribution, would be enough  to confirm the direction of solar motion as the median inverse recoil direction at the 95\% CL. Ref.~\cite{Green:2010gw} assumes the recoil directions, including their senses, can be reconstructed perfectly in 3d and the background is zero. Also 25 events in the 5 keV to 50 keV range in this case were found in Ref.~\cite{Billard:2009mf} to be necessary to determine the mean recoil direction (with a 20$^{\circ}$ uncertainty), assuming a less optimistic model for a CF$_4$ detector with a 50\% background contamination. With 100's of events we could detect a ring of maximum recoil rate around  $-{\bf V}_{\rm lab}$, which would be a secondary indicator of dark matter~\cite{Ring-like}.

After having obtained a dark matter signature, the next goal will be to study the WIMP properties, in particular its local velocity distribution.  With 100 events Ref.~\cite{Lee:2012pf} find that $V_{\rm lab}$, the magnitude of ${\bf V}_{\rm lab}$, and the WIMP velocity dispersion could be determined for a 50 GeV/$c^2$  WIMP with an uncertainty of about 100 km/s (in CF$_4$). This uncertainty is large with respect to the range of the best current estimates of  $V_{\rm lab}$ (between 180 km/s and 340 km/s; see Table~\ref{table:Vlab}) and the dispersion (usually taken to be between 170 km/s and close to 300 km/s), but such a determination would be of extreme importance. Here we introduce the idea of using aberration features of the directional recoil rate to obtain information on the local WIMP velocity distribution.

In directional detectors there will be a change in the recoil direction pattern, which we call aberration after a similar effect on the position of stars in the sky. These aberration features in a dark matter signal are due to the change in the direction of the arrival velocity of WIMPs on Earth caused by Earth's motion around the Sun. These features depend on the orbital characteristics of Earth's revolution and on the WIMP velocity distribution, and knowing the former they can be used to determine the latter.

Due to Earth's revolution, the mean recoil direction will change with a period of one year. The maximal angular annual  separation $\gamma_s$ between the mean recoil directions in a year is inversely proportional to $V_{\rm lab}$. Depending on the velocity of the rotation of the Galaxy at the position of the Sun, $\bf{V}_{\rm {GalRot}}$, this angular separation could be between 18$^\circ$ and 11$^\circ$ (for $V_{\rm {GalRot}}$  between 180 km/s and 312 km/s in the IMB model). The percentage error in the determination of the WIMP velocity is in this method the percentage error in the determination of $\gamma_s$. Thus we need a precision $\Delta\gamma_s$ of a few degrees. Using the number of events given in Refs.~\cite{Lee:2012pf} and \cite{Billard:2010jh} to determine the mean recoil direction within a few degrees, in Section 4 we estimate that a number of events between 1400 and a few thousands (between 60 kg-yr and 120 kg-yr exposure of CF$_4$) would be needed to measure the angular difference in recoil directions integrated over six or three months, which varies in the IMB model between 11$^\circ$ and 7$^\circ$  or  16$^\circ$ and 9$^\circ$, respectively.

With more than a thousand events, e.g.~more than 60 kg-yr of CF$_4$ (see Section 6), it would also be possible to detect the annual modulation of the integrated rate due to aberration. The usual non-directional annual modulation of the rate integrated over all directions depends only on the annual change in the magnitude of  ${\bf V}_{\rm lab}$. The modulation due to aberration is much larger. In Section 5 we show that the annual modulation of the rate integrated over some Galactic hemispheres, which we call Galactic Hemisphere Annual Modulations (GHAM), have much larger amplitudes than the usual non-directional modulation. The hemispheres with the largest GHAM are those divided by planes perpendicular to Earth's orbit around the Sun, since for them the full orbital velocity is in the direction of one hemisphere at one time and away from it six months later. The GHAM amplitudes as functions of energy depend strongly on the WIMP average velocity and velocity dispersion, and for the energy differential rate change sign at much lower energies than the usual non-directional annual modulation amplitude (see Section 5).

We start in Sections 2 and 3 by presenting WIMP and recoil aberration maps.

\section{Annual variation of the WIMP directional  flux}

In the IMB, WIMPs are on average at rest with respect to the Galaxy, and have a Maxwellian velocity distribution with dispersion $\sigma_v$, truncated at the escape speed $v_{\rm esc}$ (with respect to the Galaxy). Normalized to 1, the WIMP velocity distribution in the laboratory rest-frame is given by~\cite{Gondolo:2002}
\begin{equation}
f_{\rm WIMP}({\bf v})=\frac{1}{N_{\rm esc} (2\pi \sigma_v^2)^{3/2}}\exp{\left[ -\frac{({\bf v}+{\bf V}_{\rm lab})^2}{2 \sigma_v^2} \right]},
\label{VelDist}
\end{equation}
for $|{\bf v}+{\bf V}_{\rm lab}|<v_{\rm esc}$, and zero otherwise.
Here ${\bf v}$ is the WIMP velocity relative to the detector, and
\begin{equation}
N_{\rm esc}=\mathop{\rm erf}\left(\frac{v_{\rm esc}}{\sqrt{2}\sigma_v}\right)-\sqrt{\frac{2}{\pi}}\frac{v_{\rm esc}}{\sigma_v}\exp{\left[-\frac{v_{\rm esc}^2}{2\sigma_v^2} \right]}.
\label{Radon-transform}
\end{equation}
The laboratory is moving with velocity $\textbf{V}_{\rm lab}$ with respect to the Galaxy (thus  $- \textbf{V}_{\rm lab}$ is the average velocity of the WIMPs with respect to the detector). Following Ref.~\cite{Kuhlen},  which gives 100 km/s and 130 km/s as extreme estimates for the 1d velocity dispersion $\sigma_v/\sqrt{3}$, we take $\sigma_v$ either 173 km/s or 225 km/s.
A recent study by the RAVE survey using high velocity stars finds an escape speed in the range $498~{\rm {km/s}}<v_{\rm esc}<608$~km/s at 90\% confidence, with a median-likelihood value of 544 km/s~\cite{RAVE}. We use the median RAVE value, $v_{\rm esc}=544$~km/s in this paper.

  \begin{figure}[t]
\begin{center}
  \includegraphics[height=300pt]{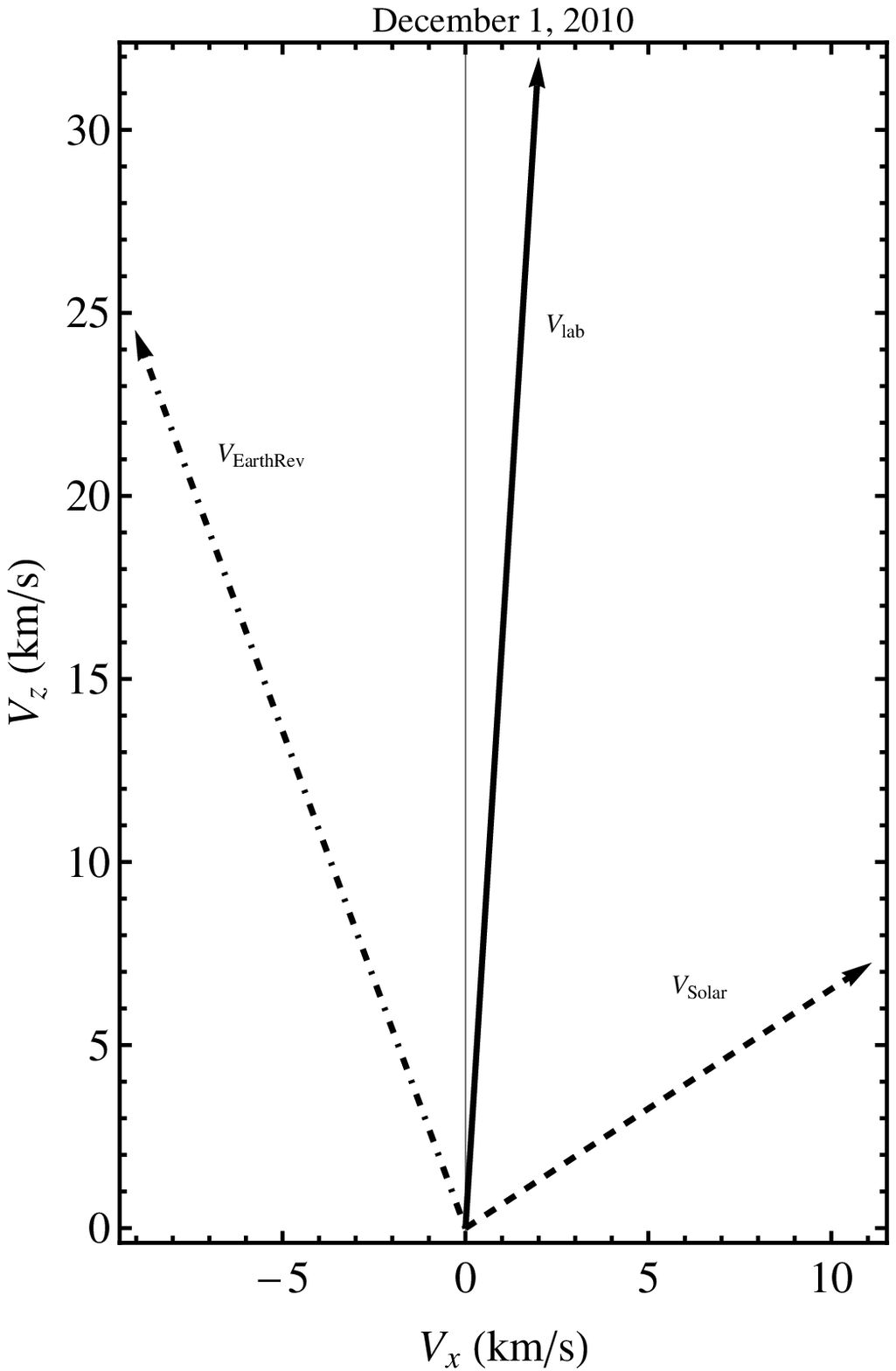}
  \includegraphics[height=300pt]{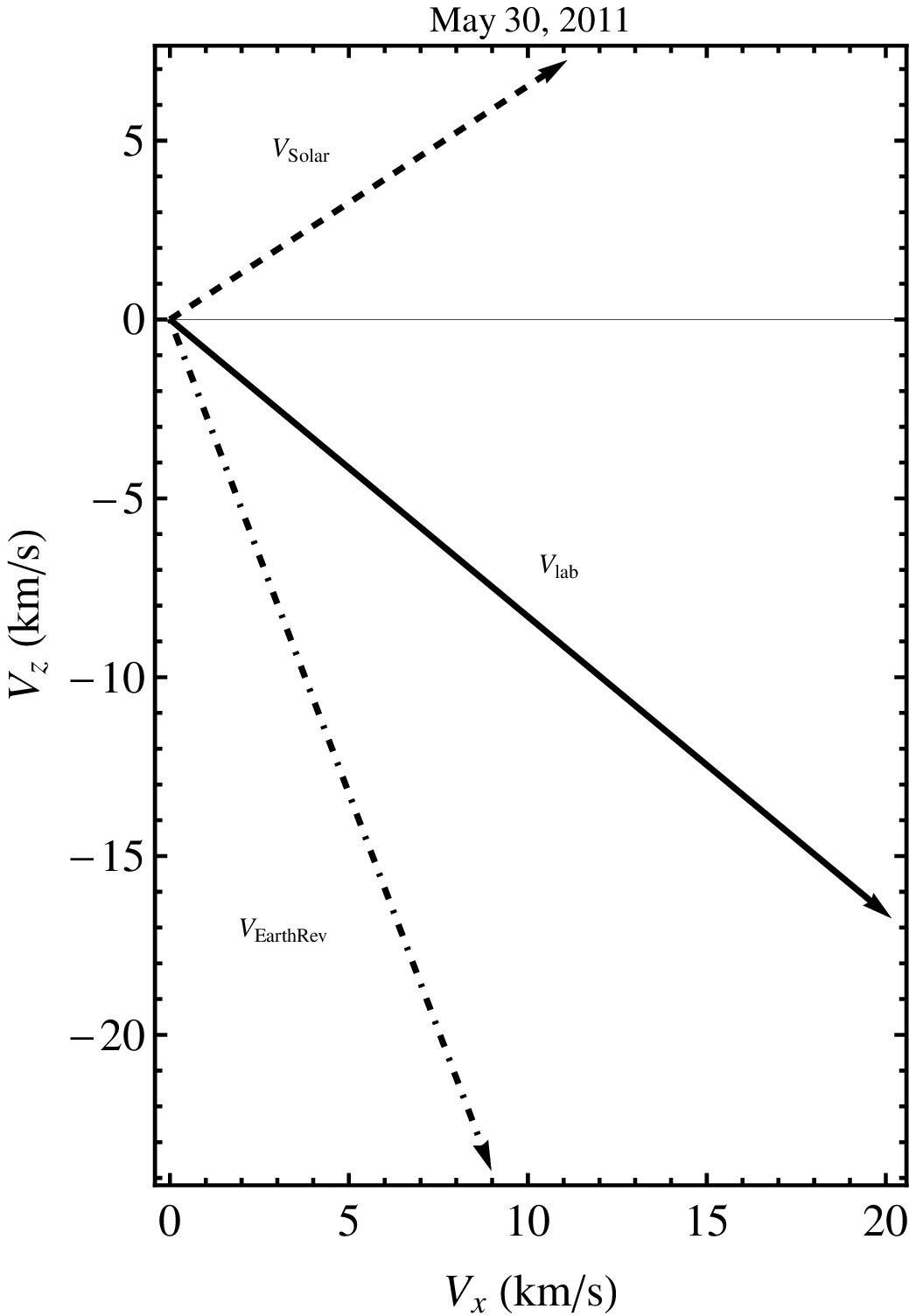}\\
  \vspace{-0.1cm}\caption{${\bf V}_{\rm {Solar}}$, ${\bf V}_{\rm {Earth Rev}}$, and ${\bf V}_{\rm {lab}}$ shown in the $x_g$-$z_g$ Galactic plane for (a) December 1 and (b) May 30. The uncertainties in $V_{\rm {Solar}}$ and $V_{\rm {Earth Rev}}$ are small enough that this result does not change (see text). Here $V_{\rm GalRot}=180$ km/s.}
  \label{Vlab-plot}
\end{center}
\end{figure}

The velocity of the lab with respect to the center of the Galaxy $\textbf{V}_{\rm lab}$ can be divided into four components: the Galactic rotation velocity ${\bf V}_{\rm {GalRot}}$ at the position of the Sun or Local Standard of Rest (LSR) velocity, Sun's peculiar velocity ${\bf V}_{\rm {Solar}}$ in the LSR, Earth's translational velocity ${\bf V}_{\rm {Earth Rev}}$ with respect to the Sun, and the  velocity of Earth's rotation around its axis ${\bf V}_{\rm {Earth Rot}}$,
\begin{equation}
{\bf V}_{{\rm lab}}= {\bf V}_{\rm {GalRot}}+  {\bf V}_{\rm {Solar}}+ {\bf V}_{\rm {Earth Rev}}+ {\bf V}_{\rm {Earth Rot}}.
\label{Vlab}
\end{equation}

The standard value of $V_{\rm {GalRot}}$ for the Standard Halo Model is 220 km/s~\cite{Kerr-1986}. As discussed in Ref.~\cite{Green:2010gw}, recent studies have found other values for  $V_{\rm {GalRot}}$. One analysis found $V_{\rm {GalRot}}=(254 \pm 16)$ km/s~\cite{Reid-2009}. Another study~\cite{Bovy-2009} found  $V_{\rm {GalRot}}=(236 \pm 11)$ km/s assuming a flat rotation curve, while Ref.~\cite{McMillan-2010} found values ranging from $V_{\rm {GalRot}}=(200 \pm 20)$ km/s to $V_{\rm {Gal Rot}}=(279 \pm 33)$ km/s depending on the model used for the rotation curve. We take $V_{\rm {Gal Rot}}=180$ km/s and 312 km/s as low and high estimates.

Due to the ellipticity of the Earth's orbit the times of maximum and minimum $V_{\rm {lab}}$ are not exactly half a year apart. These times depend on $V_{\rm {Gal Rot}}$. For $V_{\rm {Gal Rot}}=180$ km/s, $V_{\rm {lab}}$ is maximum on May 30 and minimum on December 1. For $V_{\rm {Gal Rot}}=312$ km/s, the maximum and minimum happen on June 2 and December 5, respectively. Table~\ref{table:Vlab} gives the maximum and minimum values of $V_{\rm {lab}}$ (in km/s) for our two choices of $V_{\rm GalRot}$, including the contribution of the Solar motion ${\bf V}_{\rm Solar}$.

Ref.~\cite{Schoenrich-2010} find ${\bf V}_{\rm {Solar}}=(11.1^{+0.69}_{-0.75}, 12.24^{+0.47}_{-0.47}, 7.25^{+0.37}_{-0.36})$ km/s in the Galactic reference frame, with additional systematic uncertainties $\sim (1,2,0.5)$ km/s. The uncertainty in $V_{\rm {Solar}}$ may shift the dates of the maximum and minimum of $V_{\rm {lab}}$ by only one day. Assuming $V_{\rm {Gal Rot}}=180$ km/s, and taking the low value of $V_{\rm {Solar}}$, the maximum and minimum of  $V_{\rm {lab}}$ occur on May 30 and December 2, respectively. For the high value of $V_{\rm {Solar}}$, the maximum and minimum of  $V_{\rm {lab}}$ occur on May 29 and December 1, respectively. The changed caused in the dates of the maximum and minimum $V_{\rm {lab}}$ by the uncertainty in $V_{\rm {EarthRev}}$ is less than one day.

\begin{table}[t]
\begin{center}
\begin{tabular}{lcccc}
\hline
\hline
Date & ~~~$V_{\rm {Gal Rot}}$ (km/s)~~~ & ~~~$\left|{\bf V}_{\rm {Gal Rot}}+{\bf V}_{\rm {Solar}}\right|$ (km/s)~~~ & $V_{\rm {lab}}$ (km/s)
\\
\hline
December 1 & 180 & 193 & 179.8 \\
May 30 & 180 & 193 & 208.8\\
\hline
December 5 & 312 & 324 & 310.5\\
June 2 & 312 & 324 & 340.2 \\
\hline
\hline
\end{tabular}
\caption{Maximum and minimum values of $V_{\rm lab}$ in km/s, for our two choices of the Galactic rotation speed $V_{\rm GalRot}$.}
\label{table:Vlab}
\end{center}
\end{table}

\begin{figure}[t]
\begin{center}
  \includegraphics[height=90pt]{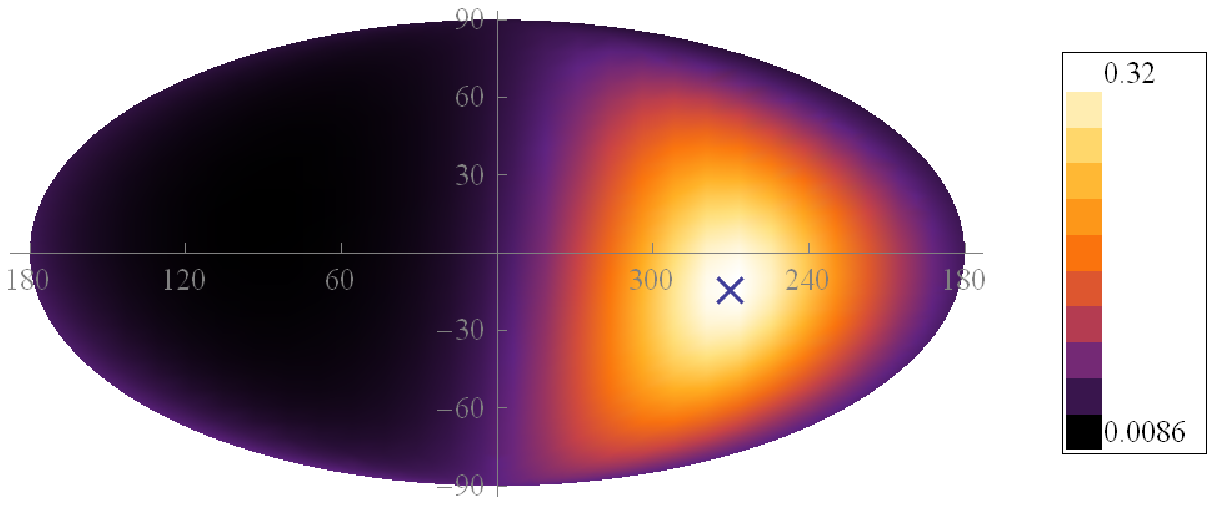}
  \includegraphics[height=90pt]{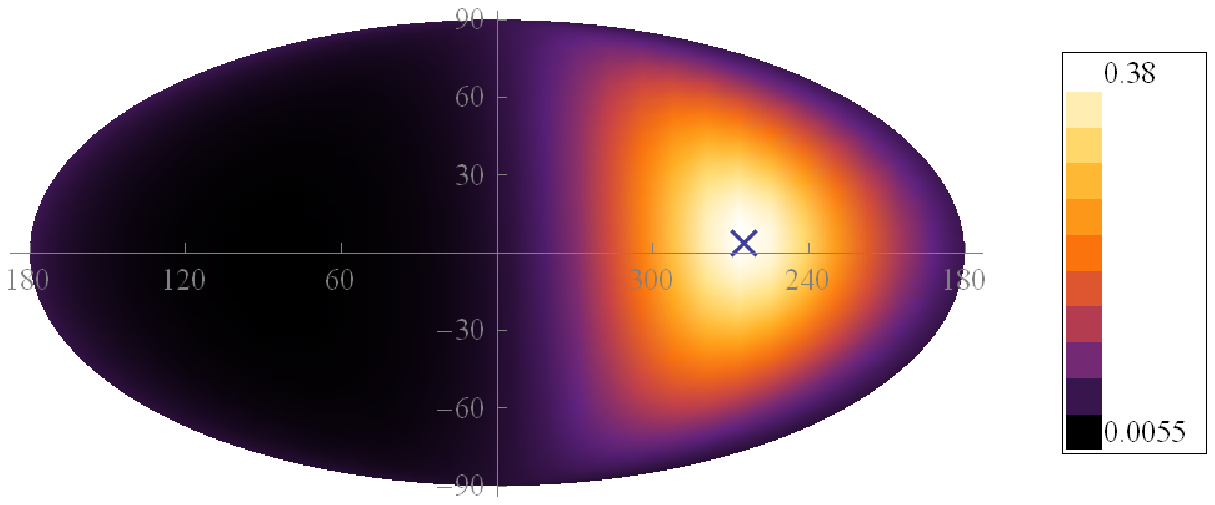}
  \includegraphics[height=90pt]{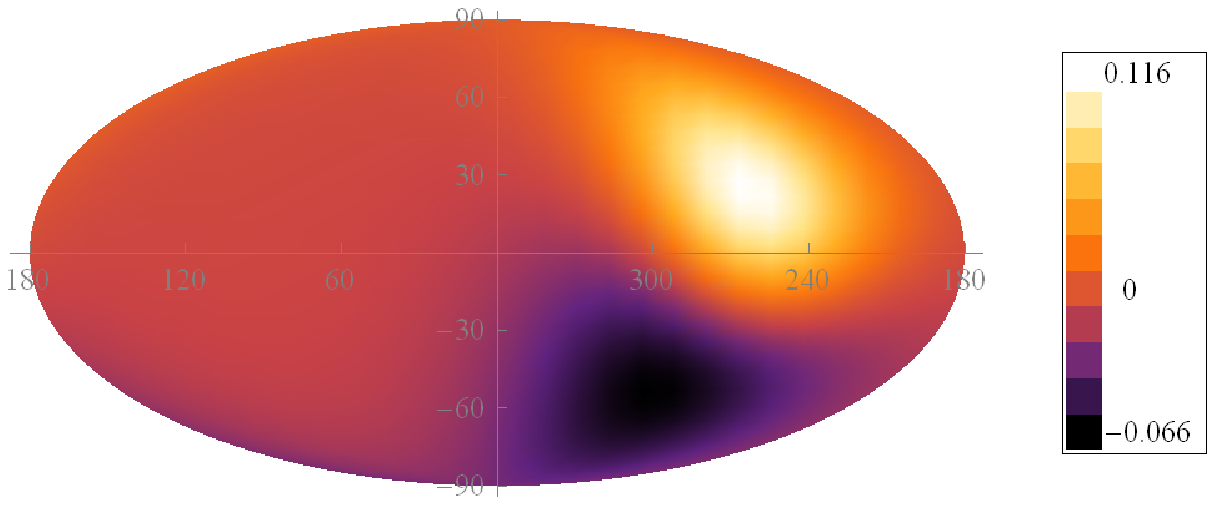}\\
  \vspace{-0.1cm}\caption{Mollweide equal-area projection maps of the celestial sphere in Galactic coordinates showing the number fraction  $F_{\rm WIMP}(\hat{\bf v},v_q)$  of $m = 100$ GeV{/$c^2$ WIMPs crossing the Earth  per unit solid angle as a function of the WIMP velocity direction $\hat{\bf v}$ with speed larger than $v_q=113$ km/s, as necessary to produce $E_R=5$ keV sulfur recoils. We assume the IMB with $v_{\rm esc}=544$ km/s, $\sigma_v=173$ km/s and $V_{\rm GalRot}=180$ km/s. (a) (Left panel) on December 1. (b) (Right panel) on May 30. Notice the direction of $-{\bf V}_{\rm lab}$ marked with  a cross. (c) Center panel: May$-$Dec difference of the number fractions. The color scale/grayscale shown in the vertical bars corresponds to  equal steps between the minimum and maximum values in units of sr$^{-1}$. The directional differential recoil rate at 5 keV in CS$_2$ corresponding to this WIMP flux is shown in Fig.~\ref{m100e5-180-173}.}}
  \label{Flux}
\end{center}
\end{figure}

\begin{figure}[t]
\begin{center}
  \includegraphics[height=90pt]{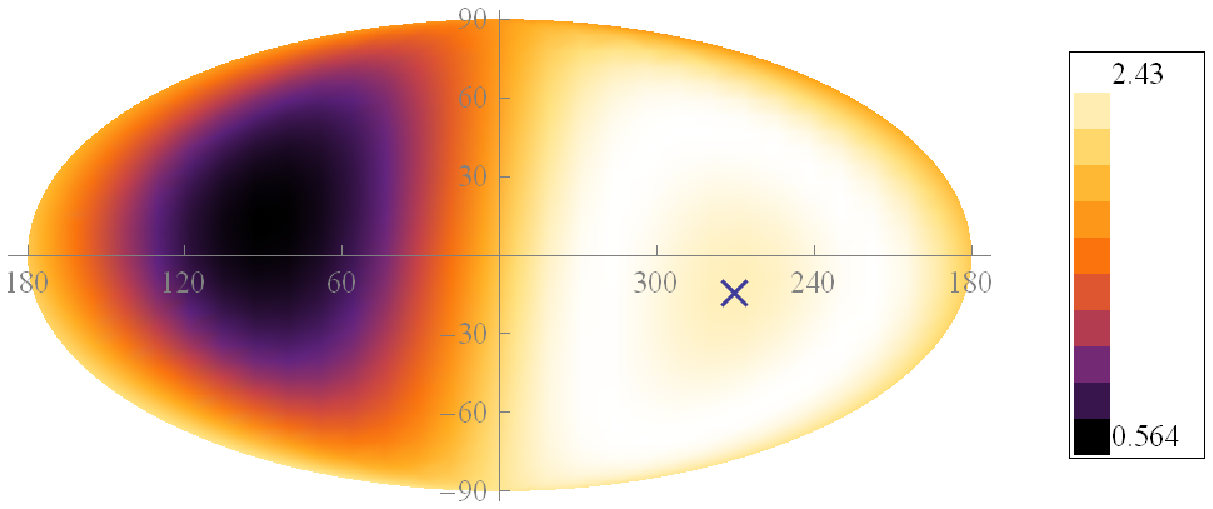}
  \includegraphics[height=90pt]{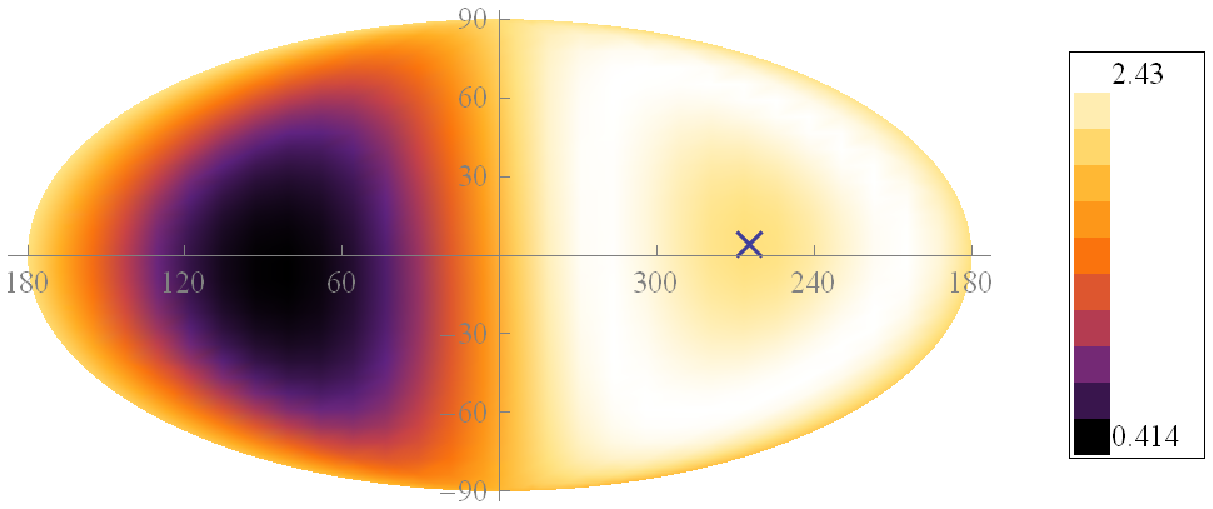}\\
  \vspace{-0.1cm}\caption{Directional differential recoil rate at $E_R=5$ keV in CS$_2$ plotted in Mollweide equal area projection maps of the celestial sphere in Galactic coordinates  for a WIMP of mass $m = 100$ GeV/$c^2$,  in the IMB with $V_{\rm GalRot}=180$ km/s, $\sigma_v=173$ km/s and  and $v_{\rm esc}=544$ km/s (same as in Fig.~\ref{Flux}) (a) on December 1, and (b) on May 30.  The direction of $-{\bf V}_{\rm lab}$ is indicated with  a cross (in the SGH in December and in the NGH in May). The color scale/grayscale shown in the vertical bars corresponds to  equal steps between the minimum and maximum values given in units of $10^{-6} \times (\rho_{0.3} \sigma_{44}/{\text{kg-day-keV-sr}})$.}
  \label{m100e5-180-173}
\end{center}
\end{figure}

In Fig.~\ref{Vlab-plot} we show the components of ${\bf V}_{\rm {Solar}}$, ${\bf V}_{\rm {Earth Rev}}$, and ${\bf V}_{\rm {lab}}$ (${\bf V}_{\rm {Earth Rot}}$ is too small to be shown) on December 1 and May 30 in the plane perpendicular to ${\bf V}_{\rm {Gal Rot}}$  (${\bf V}_{\rm {Gal Rot}}$ lies on the Galactic equatorial plane). This is the $x_g$-$z_g$ Galactic plane, where the $x_g$-axis points towards the Galactic Center and the $z_g$-axis points towards the Galactic North. Notice that the  component of ${\bf V}_{\rm {lab}}$ perpendicular to the Galactic equatorial plane  points slightly towards the North Galactic Hemisphere (NGH) in December and slightly towards the South Galactic Hemisphere (SGH) in May. This is relevant to understand Figs.~\ref{Flux} and \ref{m100e5-180-173}.

The plane of the ecliptic (on which ${\bf V}_{\rm {Earth Rev}}$ lies) is at about 60$^\circ$ of the Galactic equatorial plane, so ${\bf V}_{\rm {Earth Rev}}$ points towards the NGH for half of the year and towards the SGH for the other half and $V_{\rm Earth Rev} \simeq 30$ km/s (the average orbital speed is 29.8 km/s) is larger than $V_{\rm Solar} \simeq 18$ km/s. The component of ${\bf V}_{\rm Solar}$ in $x_g$-$z_g$ has a magnitude of approximately 13 km/s and a constant direction, while the component of  ${\bf V}_{\rm Earth Rev}$ on the same plane, has a larger magnitude on May 30 and December 1, about 25 km/s, and inverts its direction.
The result is that the component of ${\bf V}_{\rm lab}$ in the $x_g$-$z_g$ plane points North in December  and is larger in magnitude than in May, when it points South.

The uncertainties in the measurement of ${\bf V}_{\rm Solar}$ and ${\bf V}_{\rm EarthRev}$ are small and do not change this result.  Due to the uncertainty in the ecliptic longitude of the Earth's orbit minor axis, $\lambda_0=13^\circ \pm 1^\circ$,  there is an uncertainty  in $V_{\rm {Earth Rev}}$, ${\bf V}_{\rm EarthRev}=(-9.027 \pm 0.001, -15.151 \pm 0.002, 24.529 \pm 0.004)$ km/s (in the Galactic reference frame) on December 1 and ${\bf V}_{\rm EarthRev}=(8.979 \pm 0.001, 14.725 \pm 0.002, -23.798 \pm 0.004)$ km/s on May 30~\cite{Lewin-1996,Green-2003,Lang}.

  \begin{figure}[h]
\begin{center}
  \includegraphics[height=90pt]{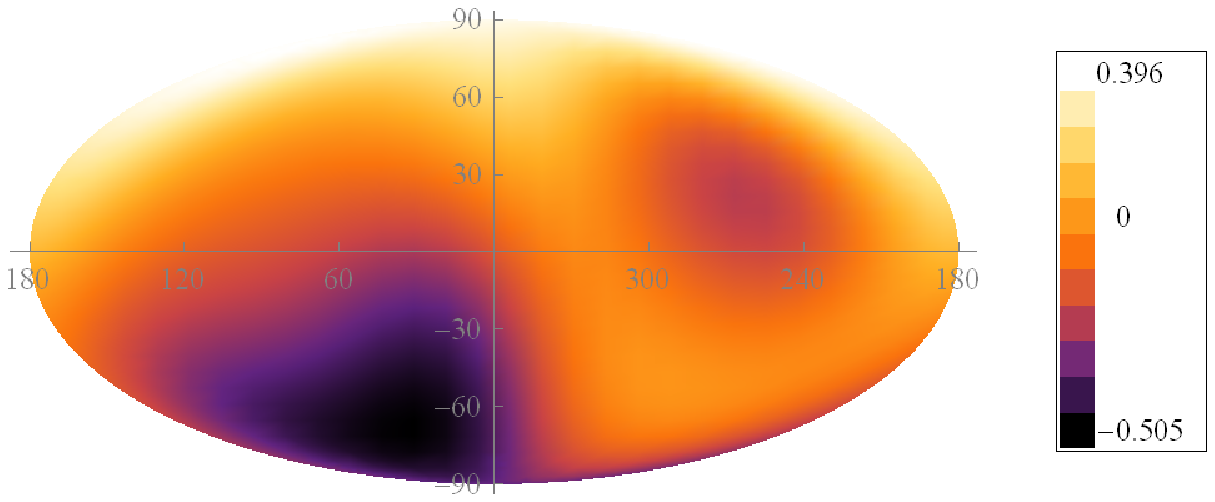}
  \includegraphics[height=90pt]{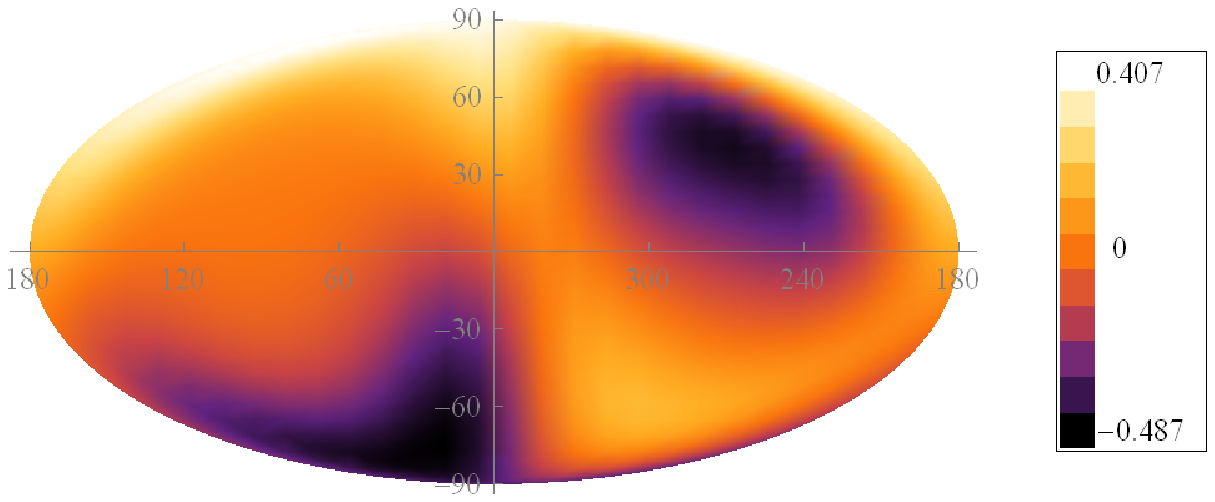}
  \includegraphics[height=90pt]{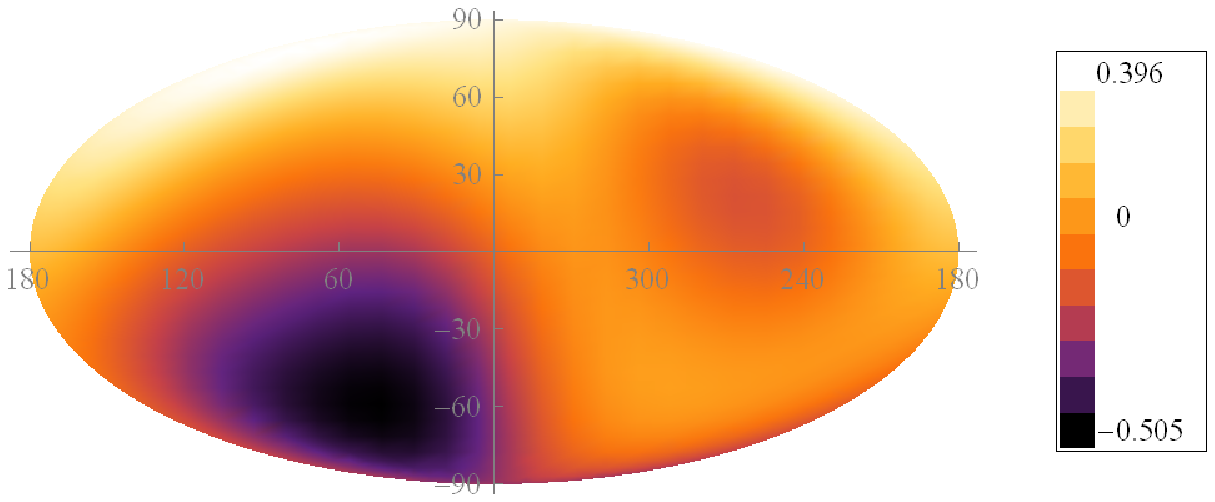}
  \includegraphics[height=90pt]{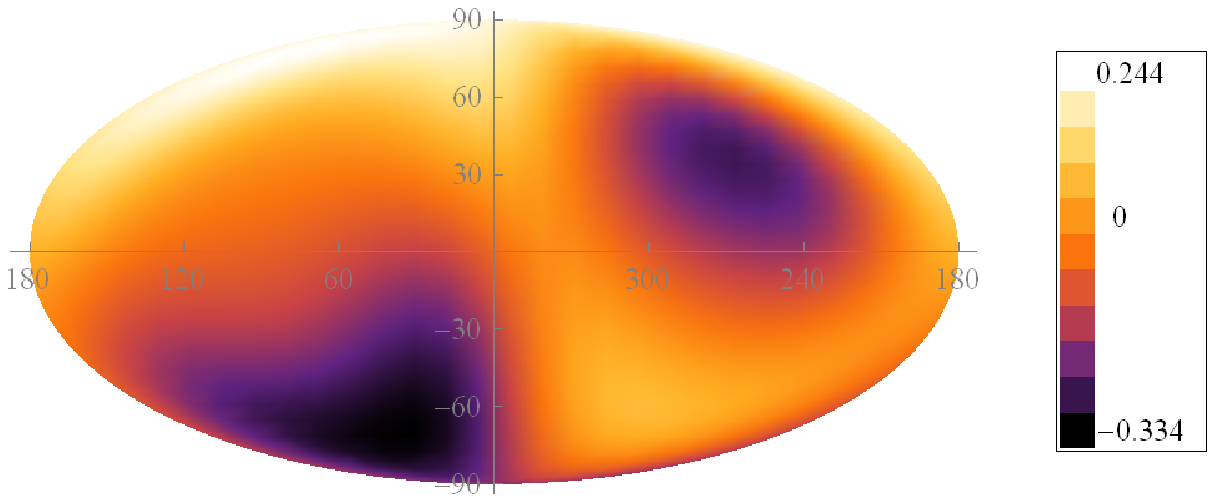}\\
  \vspace{-0.1cm}\caption{Difference of the directional differential recoil rates at $E_R=5$ keV in CS$_2$ for a WIMP of mass $m = 100$ GeV/$c^2$ and (a) May 30$-$Dec 1, $V_{\rm GalRot}=180$ km/s,  $\sigma_v=173$ km/s (top left); (b) June 2$-$Dec 5, $V_{\rm GalRot}=312$ km/s,  $\sigma_v=173$ km/s (top right); (c) May 30$-$Dec 1, $V_{\rm GalRot}=180$ km/s,  $\sigma_v=225$ km/s (bottom left); (d) June 2$-$Dec 5, $V_{\rm GalRot}=312$ km/s,  $\sigma_v=225$ km/s (bottom right). The color scale/grayscale shown in the vertical bars corresponds to  equal steps between the minimum and maximum values given in units of $10^{-6} \times (\rho_{0.3} \sigma_{44}/{\text{kg-day-keV-sr}})$.}
  \label{m100e5-Diff}
\end{center}
\end{figure}

The projection of $-{\bf V}_{\rm lab}$ onto the plane perpendicular to $-({\bf V}_{\rm GalRot}+{\bf V}_{\rm Solar})$ is maximum in March and in September (for $V_{\rm GalRot}>100$ km/s, otherwise they happen at other times). In these two months the projection points in opposite directions and $V_{\rm lab}$ happens to be the same (see Table~\ref{gammas-tab}). Thus the maximal angular separation, $\gamma_s$ between the directions of $-{\bf V}_{\rm lab}$ in a year is given by the difference between the March and September directions (for $V_{\rm GalRot}>100$ km/s).

Table~\ref{gammas-tab} gives the values of $V_{\rm {lab}}$ (in km/s) in March and September for our two choices of $V_{\rm GalRot}$, as well as the maximal angular separation $\gamma_s$ (between the March and September directions of $-{\bf V}_{\rm lab}$), and the angular separation between the mean recoil directions obtained in two three month periods (February to April and August to October) and two six month periods (January to June and July to December) centered in March and September, respectively.

\begin{table}[t]
\begin{center}
\begin{tabular}{lcc}
\hline
\hline
$V_{\rm {Gal Rot}}$ (km/s) & ~~~180~~~ & 312\\
\hline
$V_{\rm lab}$ (km/s) & 195 & 325 \\
\hline
Maximum & 17.8$^\circ$ & 10.6$^\circ$ \\
3 months & 16.2$^\circ$ & 9.0$^\circ$\\
6 months & 11.4$^\circ$ & 6.8$^\circ$\\
\hline
\hline
\end{tabular}
\caption{Values of $V_{\rm lab}$ (in km/s) in March and September, maximal angular separation (March$-$Sep) in degrees, and the angular separation between 3 month averages and 6 month averages.}
\label{gammas-tab}
\end{center}
\end{table}

The plots of  directional rate in the celestial sphere are better viewed in a planar projection. Here we use Mollweide equal-area projection maps of the celestial sphere in Galactic coordinates. The relationship between the ($x$,$y$) coordinates on a Mollweide map and the Galactic longitude and latitude ($l$,$b$) is given by (see~\cite{Gelmini-Gondolo:2001} or \cite{Ring-like})
\begin{equation}
l=\frac{-\pi x}{2\sqrt{2}\cos\theta},
\qquad
b=\arcsin{\left(\frac{2\theta+\sin(2\theta)}{\pi}\right)},
\end{equation}
where
\begin{equation}
\theta=\arcsin{(y/\sqrt{2})}.
\end{equation}

In Fig.~\ref{Flux} we show a Mollweide projection of the number fraction of 100 GeV/$c^2$ WIMPs crossing the detector per unit solid angle as a function of the WIMP velocity direction $\hat{\bf v}$ at the detector. What we plot is the number fraction of WIMPs $F_{\rm WIMP}(\hat{\bf v},v_q)$ moving in the direction $\hat{\bf v}$ with speed higher than ${v_q}$. These  are WIMPs  which can produce a recoil momentum of magnitude $q$ or recoil energy  $E_R = q^2/2M$ (5 keV in S in Fig.~\ref{Flux}) when scattering off a nucleus of mass $M$~\cite{ChanIV}
\begin{equation}
F_{\rm WIMP}(\hat{\bf v},v_q)=\int_{v_q}^{v_{\rm max}(\hat{\bf v})}{f_{\rm WIMP}({\bf v}) v^2 dv}.
\end{equation}
The upper limit of this integral is
$v_{\rm max}(\hat{\bf v})=-\hat{\bf v} \cdot {\bf V}_{\rm lab}+\sqrt{(\hat{\bf v} \cdot {\bf V}_{\rm lab})^2-{\bf V}_{\rm lab}^2+v_{\rm esc}^2}$
and the analytic expression of  $F_{\rm WIMP}(\hat{\bf v},v_q)$ is given in Eq.~13 of Ref.~\cite{ChanIV}. The maximum of $F_{\rm WIMP}(\hat{\bf v},v_q)$ happens when $\hat{\bf v} \cdot {\bf V}_{\rm lab}=-V_{\rm lab}$, i.e.~in the direction of the  average WIMP velocity $-{\bf V}_{\rm lab}$, i.e.~most WIMPs move in the direction opposite to the laboratory motion, marked by a cross in the figures.

The difference between the December and May  maps in the upper panels of  Fig.~\ref{Flux}, given in the lower panel of the figure, shows a characteristic aberration pattern with more positive numbers in the NGH and more negative ones in the SGH.  The number of WIMPs moving towards the NGH is larger in May than in December, while the opposite is true for WIMPs moving towards the SGH. This is what we would expect from the change in the direction of $-{\bf V}_{\rm lab}$ from the NGH in May to the SGH in December.

\section{The directional differential recoil spectrum}

The aberration features in directional detection depend only on the Radon transform of the WIMP velocity distribution, and not on the particular type of WIMP-nucleus interaction. The directional differential recoil spectrum as a function of the recoil momentum ${\bf q}$ is given in terms of the 3-dimensional Radon transform of the WIMP velocity distribution $\hat{ f}_{\rm lab}$~\cite{Alenazi-Gondolo:2008} as
\begin{equation}
\frac{dR}{ dE_R~d\Omega_q}= \sum_i \frac{\rho}{4\pi m \mu_i^2} C_i \hat{f}_{\rm lab}\!\left( \frac{q}{2\mu_i}, \hat{\bf q} \right) \sigma_i(q).
\label{eq: rate}
\end{equation}
Here $\rho$ is the dark matter density in the solar neighborhood and $m$ is the WIMP mass. The sum is over the nuclear species $i$  in the target, and $C_i$ and $\mu_i$ are the mass fraction and the reduced WIMP-nucleus mass for nuclide $i$, respectively. $d\Omega_q=d\phi \, d\!\cos\theta$ denotes an infinitesimal solid angle around the recoil direction $\hat{\bf q}= {\bf q}/q$, $q=|{\bf q}|$ is the magnitude of the recoil momentum, $q/2\mu=v_q$ is the minimum velocity a WIMP must have to impart a recoil momentum $q$ to the nucleus of mass $M$, or equivalently to deposit a recoil energy $E_R = q^2/2M$,   $\mu=m M/(m+M)$ is the reduced WIMP-nucleus mass,   $\sigma_i(q)$ is the WIMP-nucleus scattering cross section which can be split into spin-independent (SI) and spin-dependent (SD) parts, $\sigma_i(q)=\sigma_i^{\rm SI}(q) +\sigma_i^{\rm SD}(q)$.

The Radon transform in the laboratory frame for the truncated Maxwellian WIMP velocity distribution in Eq.~\ref{VelDist}  is~\cite{Gondolo:2002}
\begin{equation}
\hat{f}_{\rm lab}\!\left( \frac{q}{2\mu}, \hat{\bf q} \right)=\frac{1}{{N_{\rm esc}(2\pi \sigma_v^2)^{1/2}}}~{\left\{\exp{\left[-\frac{\left[ (q/2\mu) + \hat{\bf q} \cdot {\bf V}_{\rm lab}\right]^2}{2\sigma_v^2}\right]}-\exp{\left[\frac{-v_{\rm esc}^2}{2\sigma_v^2}\right]}\right\}},
\label{fhatTM}
\end{equation}
if $(q/2\mu) + \hat{\bf q} \cdot {\bf V}_{\rm lab} < v_{\rm esc}$, and zero otherwise.

The recoil momentum ${\bf q}$ is measured in a reference frame fixed to the detector. The detector  frame is at some orientation in the laboratory frame, which we define as fixed to the Earth with axes pointing to the North, the West and the Zenith. The transformation equations  between the detector frame and the laboratory frame can be conveniently written in terms of  direction cosines measurable in any experiment. This is formulated in the Appendix A of Ref.~\cite{Ring-like}, where we also give the transformations from the laboratory frame in any location on Earth to the Galactic reference frame.  The transformations in Appendix A of Ref.~\cite{Ring-like}
take into account  Earth's rotation around its axis, which is usually neglected. We use Galactic coordinates, but one could use any coordinate system to describe the aberration features.

 To give concrete examples of the GHAM we choose WIMPs with SI interactions and equal couplings to protons and neutrons interacting in  a CS$_2$ detector or with SD interactions in CF$_4$. For these WIMPs, the SI cross section in Eq.~\ref{eq: rate} is
 $\sigma_i^{\rm SI}(q)=\mu_i^2 A_i^2 \sigma_p S_i(q)/\mu_p^2$, where $\mu_p=m m_p/(m+m_p)$ is the WIMP-proton reduced mass, $\sigma_p$ is the SI WIMP-proton cross section, $A_i$ is the mass number of the nuclear species $i$, and $S_i(q)$ is the nuclear form factor, for which we use the Helm~\cite{Helm:1956} nuclear form factor normalized to 1. The SI directional differential recoil rate is therefore~\cite{Alenazi-Gondolo:2008}
\begin{equation}
\frac{dR^{\rm SI}}{dE_R~d\Omega_q} = 1.306\times10^{-3}\frac{\text{events}}{\text{kg-day-keV-sr}}\times \frac{\rho_{0.3}~ \sigma_{44}}{4\pi m \mu_p^2} \sum_i C_i \, A_i^2 \,  \, S_i(q) \hat{f}_{\rm lab}\!\left( \frac{q}{2\mu_i}, \hat{\bf q} \right).
\label{RecoilRate}
\end{equation}
Here $\rho_{0.3}$ is the dark matter density in units of 0.3 GeV/$c^2$/cm$^3$, $\sigma_{44}$ is the WIMP-proton cross section in units of $10^{-44}\;{\text{cm}}^2$, $\mu_p$ and $m$ are in GeV/$c^2$, and $\hat{ f}_{\rm lab}$ is in ${(\text{km/s})}^{-1}$.

The SD directional differential recoil rate in the proton-odd approximation in CF$_4$ is~\cite{Alenazi-Gondolo:2008}
\begin{equation}
\frac{dR^{\rm SD}}{dE_R~d\Omega_q} = 130.6~\frac{\text{events}}{\text{kg-day-keV-sr}}\times \frac{\rho_{0.3}~ \sigma_{p, 39}^{\rm SD}}{4\pi m \mu_p^2} \left[\frac{4}{3}(0.647) C_F \hat{f}_{\rm lab}\!\left( \frac{q}{2\mu_F}, \hat{\bf q} \right) \right],
\label{RecoilRate-SD}
\end{equation}
where  $\sigma_{p, 39}^{\rm SD}$ is the spin-dependent cross section off a proton in units of $10^{-39}\;{\text{cm}}^2$ (consistent with experimental bounds from direct detection and neutrino experiments~\cite{CrossSection}), and $C_F=4M_F/(4M_F+M_C)$ is the mass fraction of F in CF$_4$. Note that while the F nucleus has spin $\frac{1}{2}$, the C nucleus has no spin.

The standard value of the local dark matter density is $\rho=0.3$ GeV/$c^2$/cm$^3$, and this is what we use here. However, one should keep in mind the  large uncertainties in this parameter. Recent astronomical constraints are consistent with 0.2 GeV/$c^2$/cm$^3$ $<\rho<0.4$ GeV/$c^2$/cm$^3$ for a spherical dark matter halo profile, and up to 20\% larger for non-spherical haloes~\cite{Weber:2010}. Using a halo model independent method, Ref.~\cite{Salucci:2010} finds $\rho=0.43 \pm 0.11 \pm 0.10$ GeV/$c^2$/cm$^3$, with uncertainties from two different sources. Ref.~\cite{Pato:2010} finds the density for a specific simulated galaxy resembling the Milky Way is 21\% larger than the mean value of $\rho=0.39$ GeV/$c^2$/cm$^3$ obtained in a previous study~\cite{Catena:2010} in which spherical symmetry was assumed.

Fig.~\ref{m100e5-180-173} shows the directional  differential recoil rate (given in Eq.~\ref{RecoilRate}) in CS$_2$ at $E_R=5$ keV on December 1 (left panel (a)), and  May 30 (right panel (b)), the times at which $V_{\rm lab}$ is maximum or minimum, plotted in Mollweide maps of the recoil direction $\bf{\hat{q}}$ in Galactic coordinates  for a WIMP of mass $m = 100$ GeV/$c^2$. The  IMB values in this figure are $V_{\rm GalRot}=180$ km/s, $\sigma_v=173$ km/s, and $v_{\rm esc}=544$ km/s. The direction of the average WIMP velocity with respect to the detector, $-{\bf V}_{\rm lab}$, is indicated on the maps with a cross. The difference between the two Fig.~\ref{m100e5-180-173} maps is shown in the upper left panel of Fig.~\ref{m100e5-Diff} (in units of $10^{-6} \times (\rho_{0.3} \sigma_{44}/{\text{kg-day-keV-sr}})$) together with similar plots obtained by changing only  $V_{\rm lab}$ (i.e.~$V_{\rm GalRot}$) and  $\sigma_v$ to their maximum and minimum values.

As seen in the four panels of Fig.~\ref{m100e5-Diff}, increasing the value of $V_{\rm GalRot}$ results in a larger anisotropy in the aberration features. This is expected since the anisotropy in the flux of WIMPs arriving on Earth and thus in the differential recoil rate increases with $V_{\rm GalRot}$ (the flux would be isotropic if $V_{\rm GalRot}=0$). On the other hand, increasing $\sigma_v$ makes the difference maps more isotropic, also as expected ($\sigma_v=0$ would mean that all WIMPs arrive from only one direction). The most anisotropic aberration features (for the $V_{\rm GalRot}$ and $\sigma_v$ values that we considered) is shown in the upper right panel of Fig.~\ref{m100e5-Diff} in which $V_{\rm GalRot}$ is the largest and $\sigma_v$ the smallest, 312 km/s and 173 km/s, respectively. Notice that the right panels of Fig.~\ref{m100e5-Diff} (with $V_{\rm GalRot}=312$ km/s) show a more negative rate difference (darker region) in the NGH compared to the left panels. This is due to the higher visibility of the ring-like feature in the recoil maps of June and December for larger $V_{\rm GalRot}$ (see Figs.~5.b and 6 of Ref.~\cite{Ring-like} in which the ring is more visible for $V_{\rm GalRot}=312$ km/s). Changing the value of $v_{\rm esc}$ does not change the maps significantly, and thus we show the results only for one.

\begin{figure}[t]
\begin{center}
  \includegraphics[height=90pt]{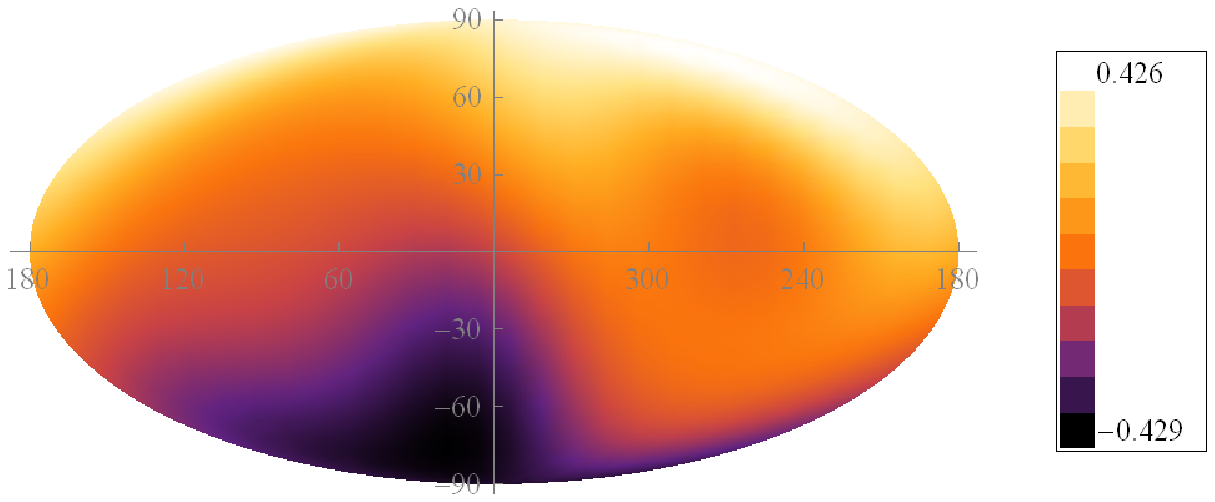}
  \includegraphics[height=90pt]{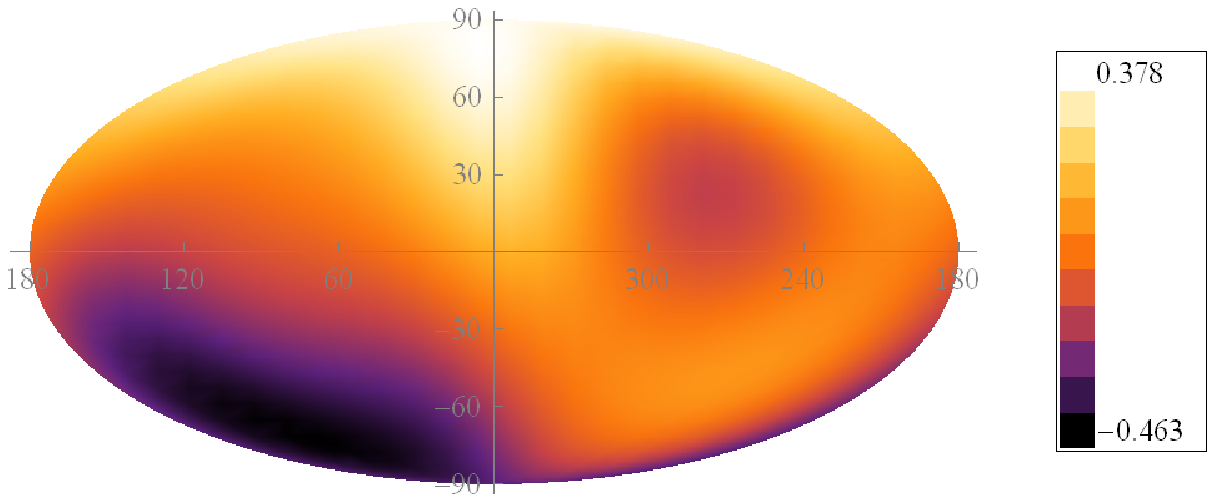}\\
  \includegraphics[height=90pt]{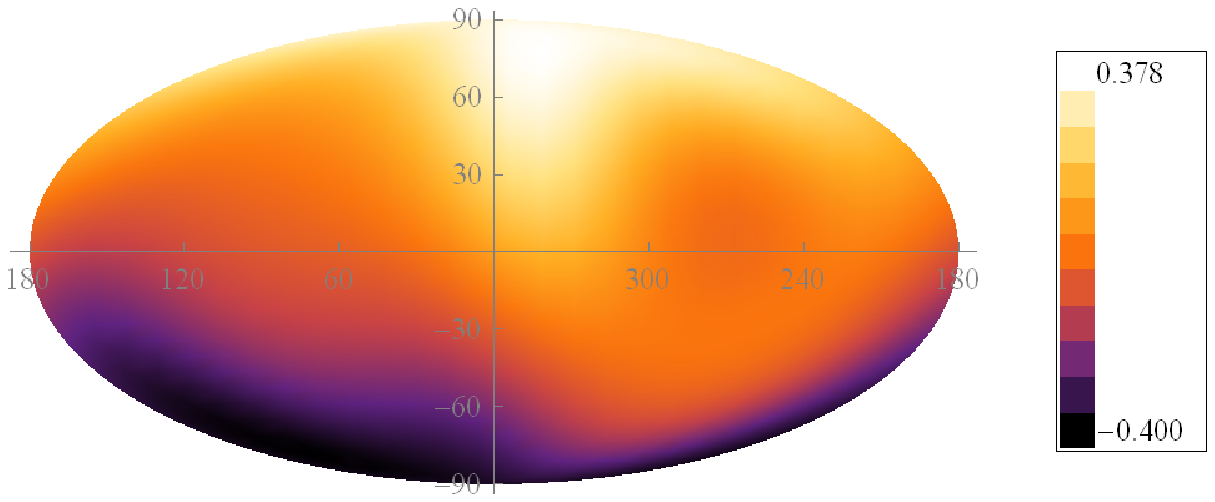}
  \vspace{-0.1cm}\caption{Same parameters as in Figs.~\ref{m100e5-180-173} and \ref{m100e5-Diff}.a. (a) (Left) (May 30)$-$(Dec 1) difference of the directional differential recoil rates averaged over an energy interval between 5 keV and 20 keV, (b) (right) difference of the time averaged directional differential recoil rates over three months at 5 keV, and (c) (center) difference of the directional differential recoil rates averaged over three months and over the 5 keV to 20 keV energy interval. In (b) and (c) the average rate of December, January, and February is subtracted from the average rate of June, July, and August.}
  \label{m100-180-173-AvgE-Avgt}
\end{center}
\end{figure}

Examining the four panels of Fig.~\ref{m100e5-Diff}, we can see that the positive directional differential rate differences (lighter regions) are mostly in the NGH and the negative rate differences (darker regions) are mostly in the SGH. This effect persists in the energy and time integrated rates. Using the same parameters of Fig.~\ref{m100e5-180-173}, Fig.~\ref{m100-180-173-AvgE-Avgt}.a shows the December and May difference in the rates averaged over the 5 keV to 20 keV energy range, Fig.~\ref{m100-180-173-AvgE-Avgt}.b shows the difference in the directional differential rates averaged over three months (December, January, and February period with respect to the  June, July, and August period), and Fig.~\ref{m100-180-173-AvgE-Avgt}.c shows the difference in the time averaged (over the same three months) and energy averaged (in the 5 keV  to 20 keV range) of the same rates. Clearly, the aberration features persist in the three panels of Fig.~\ref{m100-180-173-AvgE-Avgt}.

\begin{figure}[t]
\begin{center}
  \includegraphics[height=90pt]{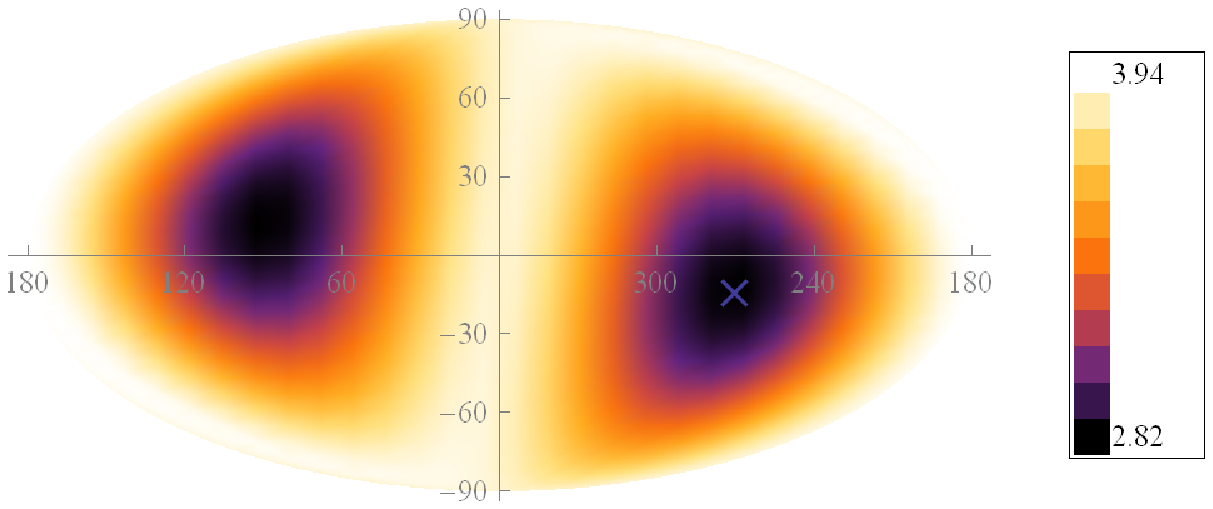}
  \includegraphics[height=90pt]{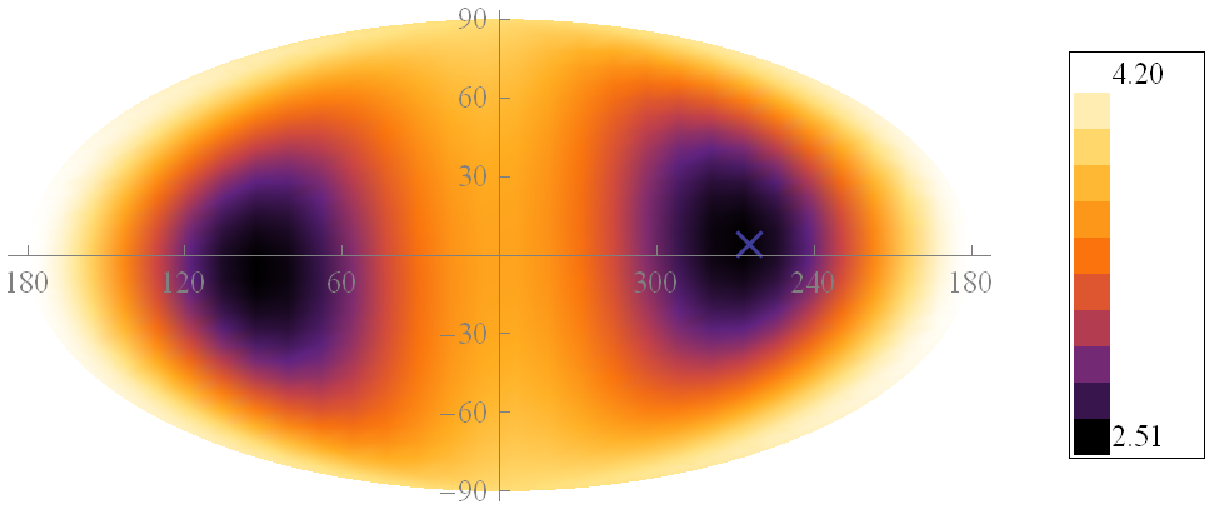}
  \includegraphics[height=90pt]{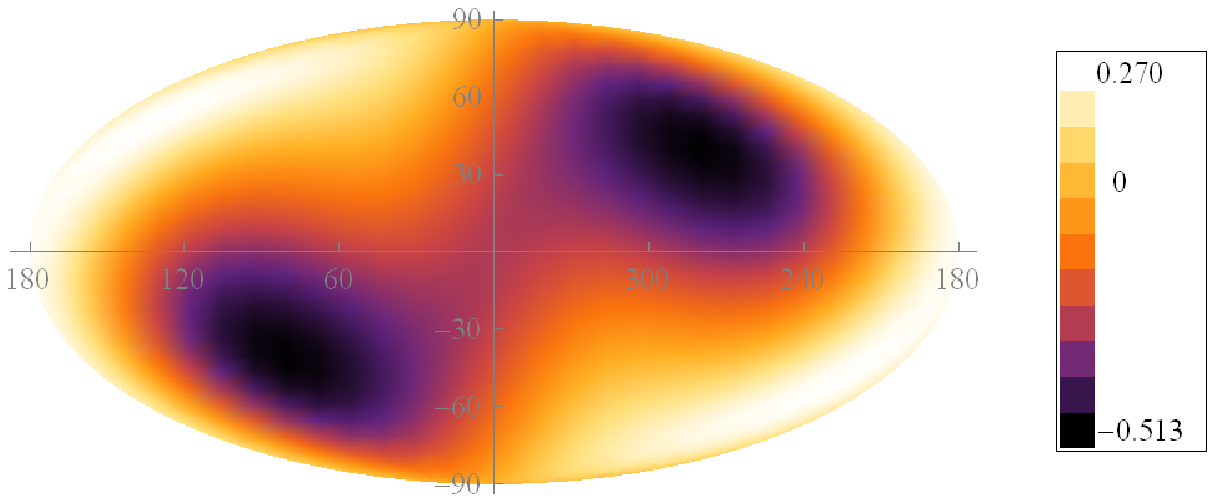}\\
  \vspace{-0.1cm}\caption{(a) and (b) same as Figs.~\ref{m100e5-180-173}.a and \ref{m100e5-180-173}.b and (c) same as Fig.~\ref{m100e5-Diff}.a but without sense discrimination.}
  \label{m100e5-180-173-NoSense}
\end{center}
\end{figure}

In experiments that are not able to distinguish the sense of a nuclear recoil (such as NEWAGE), the directional differential recoil rate detected in opposite Galactic quadrants are identical. Fig.~\ref{m100e5-180-173-NoSense} shows the directional recoil rate without sense discrimination  in CS$_2$ at $E_R=5$ keV plotted in Mollweide maps of the recoil direction $\bf{\hat{q}}$ in Galactic coordinates  for a WIMP of mass $m = 100$ GeV/$c^2$, and assuming the IMB with $v_{\rm esc}=544$ km/s, $V_{\rm GalRot}=180$ km/s and $\sigma_v=173$ km/s. The directional differential rate is given on December 1 in the left panel (a), and on May 30 in the right panel (b). The center  panel (c) gives the difference in the directional differential rate between May and December, in units of $10^{-6} \times (\rho_{0.3} \sigma_{44}/{\text{kg-day-keV-sr}})$.
Having sense discrimination in 3d makes the detection of the aberration features easier and gives the largest annual modulation amplitudes. Thus in the rest of this paper we give the results assuming the directional detector has 3d capability of reconstruction of recoil directions, including their senses.

Even with sense discrimination, the aberration patterns require a very large number of events to be observed. We concentrate in the following sections on aberration features easier to detect.

\section{Maximal angular variation of the mean recoil direction}

\begin{figure}[t]
\begin{center}
  \includegraphics[height=200pt]{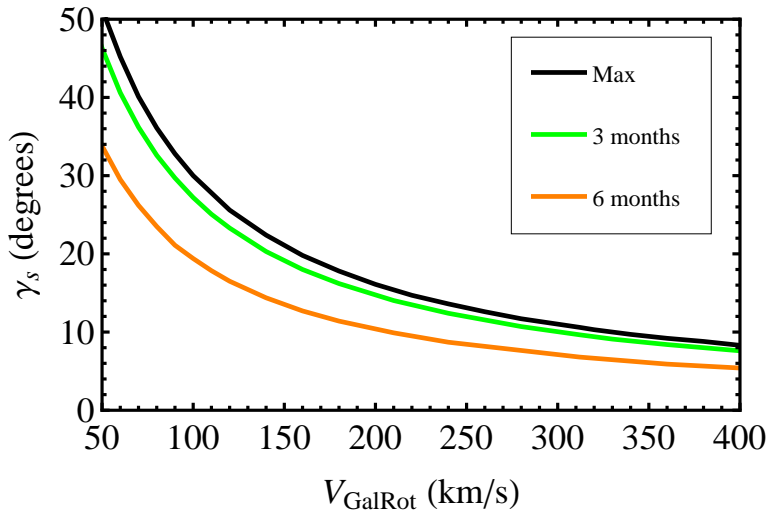}\\
  \vspace{-0.1cm}\caption{Maximal angular separation $\gamma_s$ between the $-{\bf V}_{\rm lab}$ directions in a year (black line), and between the mean recoil directions obtained in two three month periods (green line) and two six month periods (orange line), plotted as a function of $V_{\rm GalRot}$.}
  \label{AngularSep}
\end{center}
\end{figure}

The largest uncertainty in ${\bf V}_{\rm lab}$ is due to our uncertain knowledge of the velocity of the rotation of the Galaxy at the location of the Sun. Thus we can use the change in the average velocity of WIMPs with respect to Earth in a year to measure $V_{\rm GalRot}$. A simple and rather accurate evaluation of the maximal angular separation $\gamma_s$ between the directions of $-{\bf V}_{\rm lab}$ in a year is
\begin{equation}
\gamma_s \simeq \frac{2 V_{\rm EarthRev} \sin 60^\circ}{V_{\rm GalRot}} \simeq 16.5^\circ \left(\frac{180~{\rm km/s}}{V_{\rm GalRot}} \right),
\label{gammas}
\end{equation}
which gives from 16.5$^\circ$ to 9.5$^\circ$ for $V_{\rm GalRot}$ between 180 km/s and 312 km/s. The maximal angular separation, $\gamma_s$ is the difference between the $-{\bf V}_{\rm lab}$ directions in March and September. This angle $\gamma_s$ in degrees is plotted in Fig.~\ref{AngularSep}, together with the angular separations between the mean recoil directions obtained in two six month periods or two three month periods centered in March and September, respectively. $\gamma_s$ goes between 16.2$^\circ$ and 9.0$^\circ$ for the three months averaged directions, and 11.4$^\circ$ and 6.8$^\circ$ for the directions averaged over six months (see Table~\ref{gammas-tab}). The error $\Delta V_{\rm GalRot}$ in the determination of $V_{\rm GalRot}$ using $\gamma_s$ can be read off of Fig.~\ref{AngularSep} (see below). A simple estimate, $\gamma_s \sim V_{\rm GalRot}^{-1}$ indicates that  $\Delta V_{\rm GalRot}/V_{\rm GalRot} \simeq \Delta\gamma_s/\gamma_s$. Thus we need $\Delta\gamma_s$ to be a few degrees at most.

Refs.~\cite{Billard:2010jh} and \cite{Green&Morgan:2010} give estimates of the number of events needed to measure the mean recoil direction with an error of a few degrees in two different cases. Ref.~\cite{Billard:2010jh} finds that with an exposure of 30 kg-yr in CF$_4$ the mean recoil direction could be determined with an error of 2.5$^\circ$. For a 50 GeV/$c^2$ WIMP with SD interactions and $\sigma_{p}^{\rm SD}=10^{-39}\;{\text{cm}}^2$ and recoils with energy between 5 and 50 keV, as assumed in Ref.~\cite{Billard:2010jh}, 30 kg-yr corresponds to 660 events in this energy range. In Ref.~\cite{Billard:2010jh}, $V_{\rm lab}$ is fixed to 220 km/s, however it is reasonable to assume that with a variable $V_{\rm lab}$ the mean recoil direction would be determined with a similar error of a few degrees. If the data would be split in two sets (January to June and July to December), the six months averaged directional separation $\gamma_s$ could be determined with double the number of events and exposure, i.e.~1400 and 60 kg-yr. The three months averaged $\gamma_s$  would require four times the total number of events and exposure, i.e.~2800 and 120 kg-yr. Assuming an error of $\Delta\gamma_s \simeq \sqrt{2} \times 2.5^\circ \simeq 3.5^\circ$, it can be seen from Fig.~\ref{AngularSep} that the measured $V_{\rm GalRot}$ would be $180^{+60}_{-30}$ km/s or $312^{+190}_{-70}$ km/s using the three month averaged (green) curve, and $180^{+90}_{-40}$ km/s or $312^{+290}_{-110}$ km/s using the six month averaged (orange) curve.

Ref.~\cite{Green&Morgan:2010} shows in its Fig.~1 that more than $10^3$ events above 20 keV would be necessary in S to determine the mean recoil direction with an error smaller than 5$^\circ$ (in this case $m=100$ GeV/$c^2$).

Thus, both Refs.~\cite{Billard:2010jh} and \cite{Green&Morgan:2010} imply that a few thousand events would be necessary to measure $\gamma_s$. With this number of events it might be possible to detect the annual modulation of the rate integrated over Galactic hemispheres to which we devote the following two sections.

We would like to point out that $V_{\rm GalRot}$ could also be obtained from the annual average value of the projections of recoil momenta onto planes containing the mean recoil direction. One such plane could be chosen to be perpendicular to a vector at 90$^\circ$ from both Earth's orbital plane and the mean recoil direction. Another could be the plane perpendicular to this one and also containing the mean recoil direction. The average of the projections of the recoil momenta onto these two planes should be $\simeq V_{\rm GalRot}$ while the average of the projections onto the plane perpendicular to the two just mentioned would be zero. We do not develop this method any further for the time being, because we think it would not require a smaller number of events than the previous method.

\section{Annual modulation integrated over direction}

The annual modulation  of the differential rate is due to the change in the magnitude of $V_{\rm lab}$ during a year~\cite{Drukier}. At high energies, the differential rate is maximum and minimum respectively when $V_{\rm lab}$ is maximum and minimum, but at low energies the phase is inverted and the differential rate is maximum when  $V_{\rm lab}$ is minimum and vice versa.

In order to compute the annual modulation amplitude of the energy differential rate, we integrate the directional differential recoil rate over direction, and define
\begin{equation}
\Delta\left(\frac{dR}{dE_R}\right)=\frac{dR_{\textrm{max}}}{dE_R}-\frac{dR_{\textrm{min}}}{dE_R}=\int{\left(\frac{dR_{\textrm{max}}}{dE_R~d\Omega_q}-\frac{dR_{\textrm{min}}}{dE_R~d\Omega_q}\right)} d\Omega_q,
\label{AnnualMod}
\end{equation}
where the subscripts ``max'' and ``min'' refer to the maximum and minimum differential rates during a year. In the plots we show the amplitude at high energies as positive (thus it becomes negative at low energies).

In directional detection, we can integrate over the recoil directions pointing to half of the sky and define the annual modulation of the rate so obtained. This is what we call the Galactic Hemisphere Annual Modulation (GHAM). Choosing well the hemisphere, the GHAM amplitude is larger than the usual annual modulation amplitude, thus easier to detect.

\subsection{Maximum GHAM amplitudes}

As mentioned earlier, in the Mollweide figures showing the difference in the directional differential recoil rates between the two dates at which $V_{\rm lab}$ is maximum and minimum, one can see that the positive rate differences are mostly in the NGH and the negative rate differences are mostly in the SGH. This happens because the plane of the Earth's orbit around the Sun is at about 60$^\circ$ of the Galactic equatorial plane. We can compute the annual modulation amplitude for any Galactic hemisphere, but those with the largest GHAM are divided by planes that are perpendicular to the Earth's orbit around the Sun, since then the total orbital velocity is in the direction of one hemisphere at some time and away from it half a year later.

Any hemisphere can be defined by a vector perpendicular to a plane dividing the sky in two, and pointing towards the particular hemisphere. We call this vector the hemisphere's pole. Let us consider the Galactic hemisphere with the vector ${\bf V}_{\rm {Earth Rev}}$ at a particular time of the year, the date D1, as its pole. The maximum differential rate for this hemisphere happens on the date D1, and the minimum rate happens on the date D2 when ${\bf V}_{\rm {Earth Rev}}$ points in the opposite direction.

The two hemispheres with the maximum GHAM amplitudes at high and at low energies have as their poles ${\bf V}_{\rm {Earth Rev}}$ on the dates at which $V_{\rm lab}$ is maximum and minimum, respectively.

  \begin{figure}[t]
\begin{center}
  \includegraphics[height=130pt]{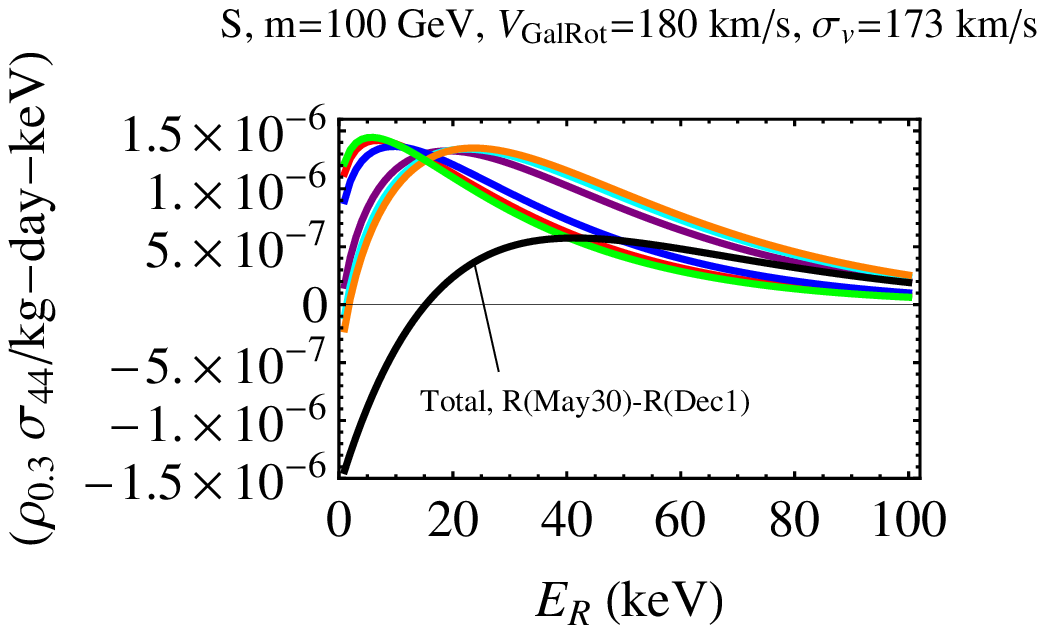}
  \includegraphics[height=130pt]{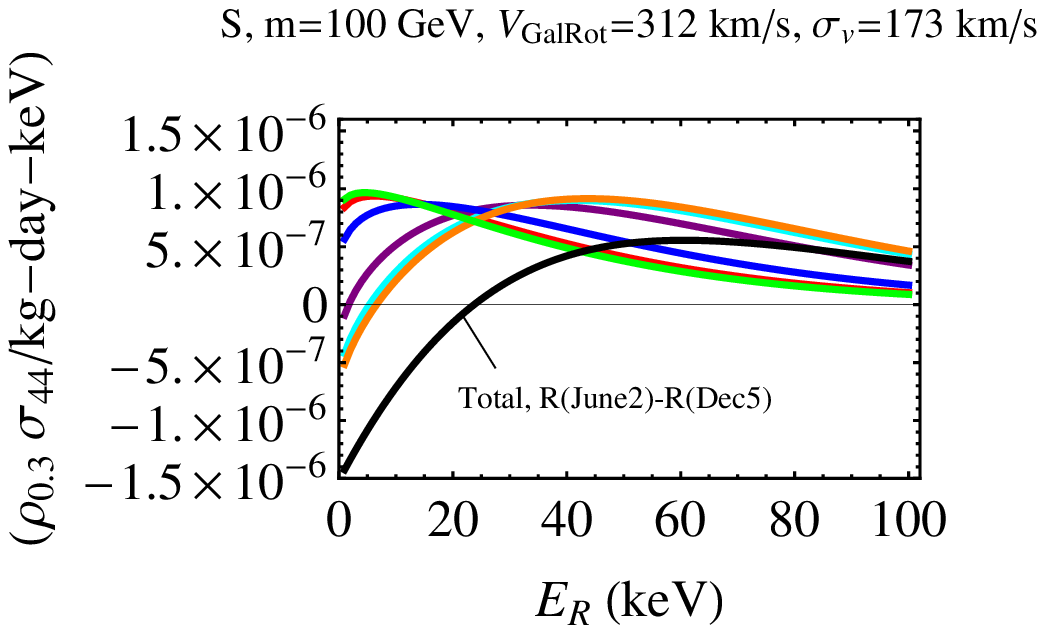}
  \includegraphics[height=130pt]{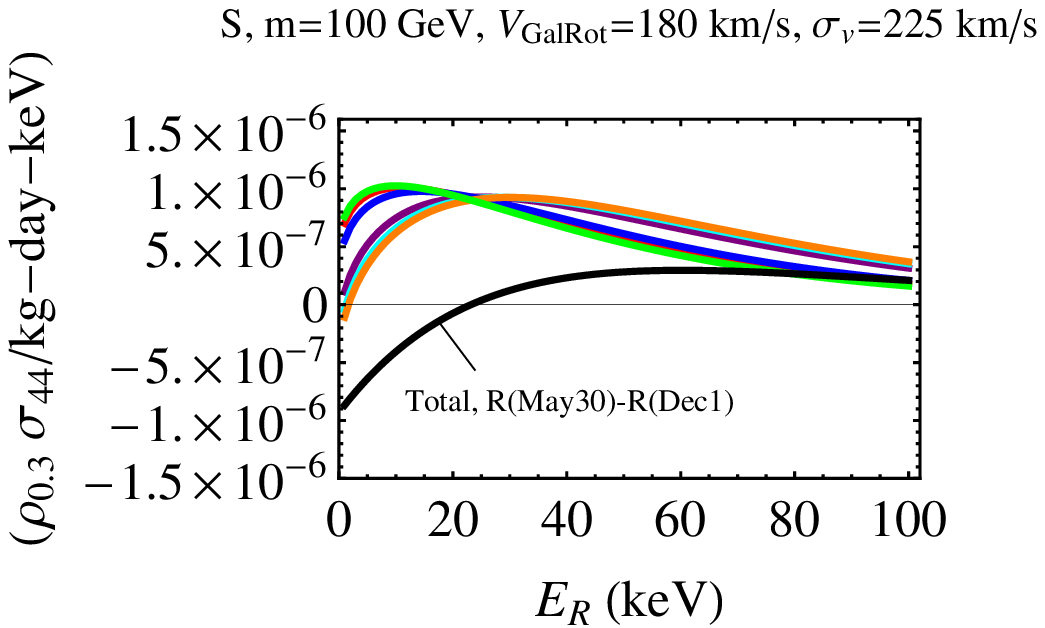}
  \includegraphics[height=130pt]{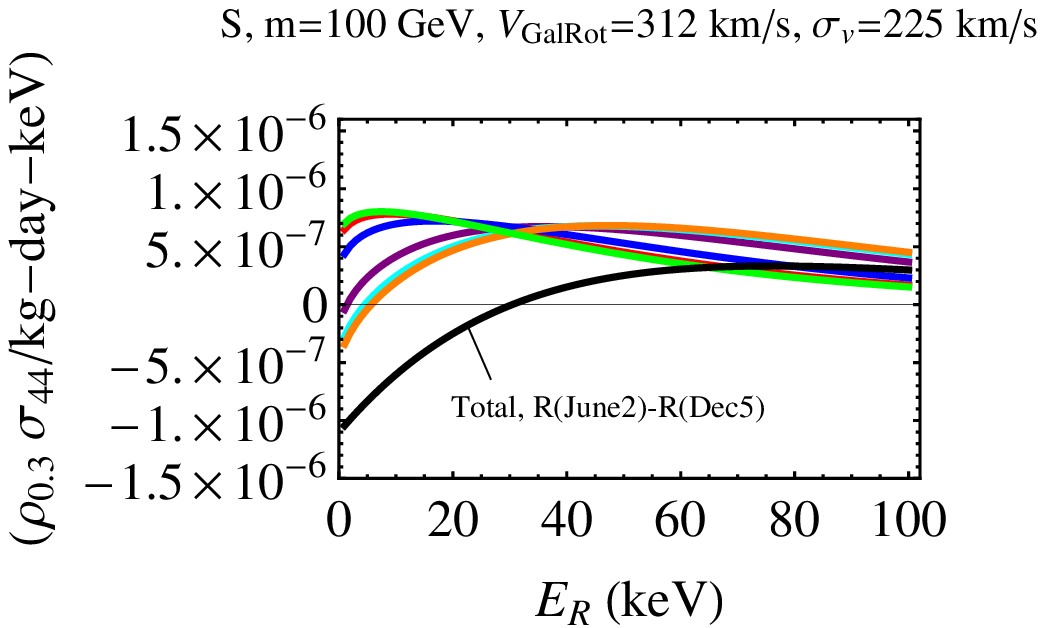}\\
  \vspace{-0.1cm}\caption{Annual modulation amplitude (Eq.~\ref{AnnualMod}) of the S energy differential recoil rate integrated over different Galactic hemispheres (colored lines) and the whole sky (black line) as a function of recoil energy $E_R$ for $m=100$ GeV/$c^2$, $v_{\rm esc}=544$ km/s and (a) $V_{\rm GalRot}=180$ km/s and  $\sigma_v=173$ km/s (top left), (b) $V_{\rm GalRot}=312$ km/s and $\sigma_v=173$ km/s (top right), (c) $V_{\rm GalRot}=180$ km/s and $\sigma_v=225$ km/s (bottom left), and (d) $V_{\rm GalRot}=312$ km/s and $\sigma_v=225$ km/s (bottom right). The red, purple, cyan, and blue curves are for different dates (rate at D1 minus rate at D2) in a year separated by three months: D1 January 1, April 1, July 1, and October 1, respectively, and D2 July 2, October 4, December 31, and March 29, respectively. The green and orange lines show the GHAM with largest amplitude at low and high energies, respectively, and they correspond to complimentary hemispheres. For the green lines, the D1$-$D2 dates are Dec 1$-$May 30 in (a) and (c), and Dec 5$-$June 2 in (b) and (d). For the orange lines the D1 and D2 dates are exchanged.}
  \label{Rate-GHAM-m100}
\end{center}
\end{figure}

  \begin{figure}[t]
\begin{center}
  \includegraphics[height=130pt]{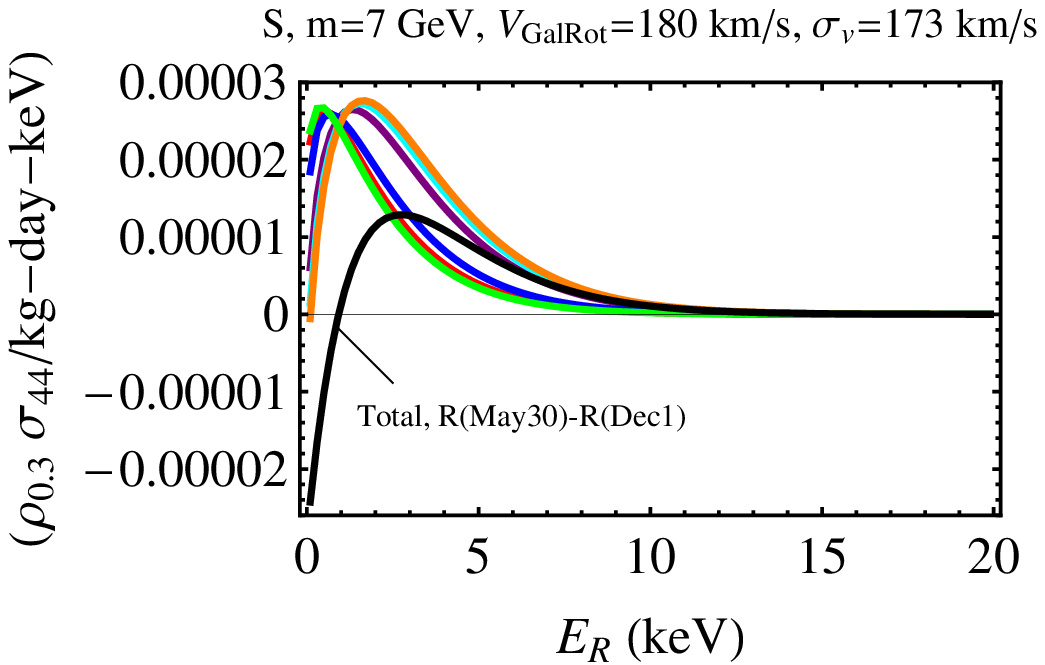}
  \includegraphics[height=130pt]{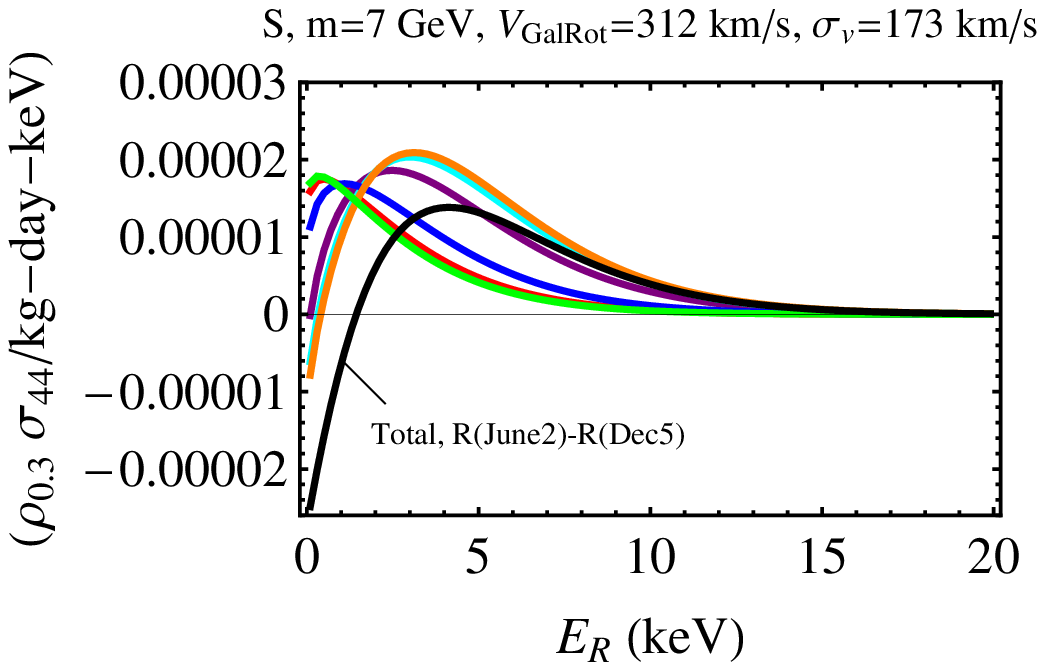}
  \includegraphics[height=130pt]{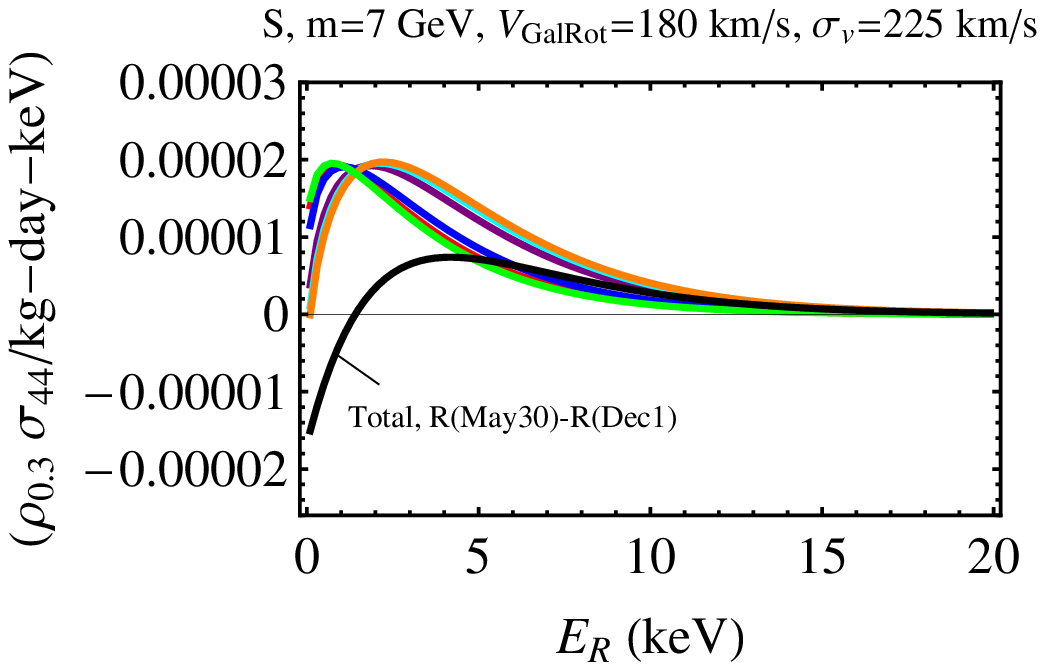}
  \includegraphics[height=130pt]{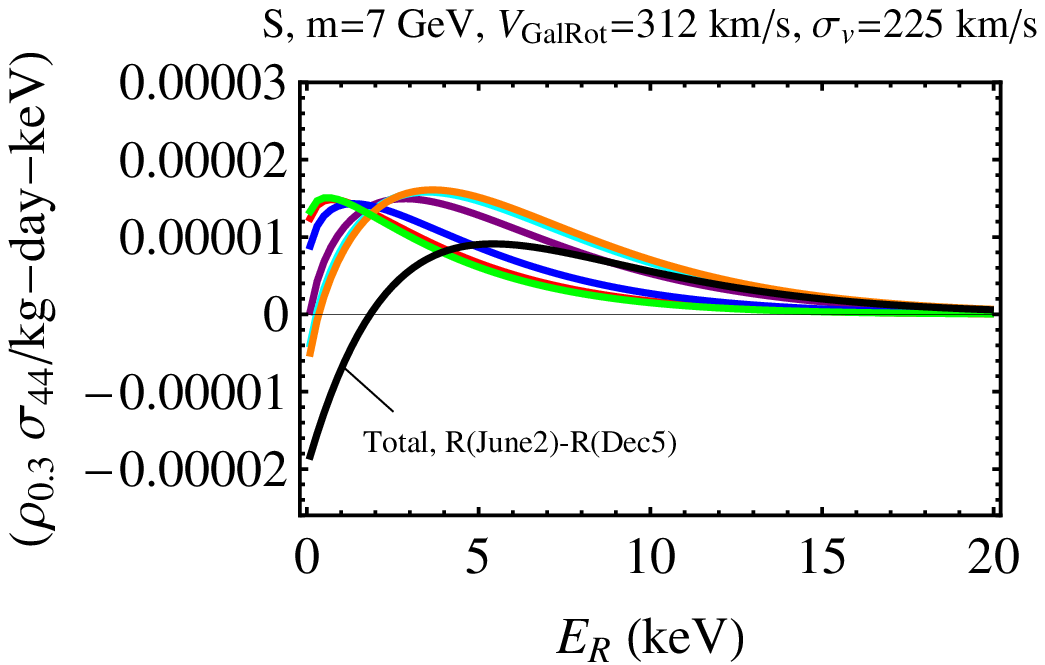}\\
  \vspace{-0.1cm}\caption{Same as Fig.~\ref{Rate-GHAM-m100} but for $m=7$ GeV/$c^2$.}
  \label{Rate-GHAM-m7}
\end{center}
\end{figure}
Figs.~\ref{Rate-GHAM-m100} and \ref{Rate-GHAM-m7} show the annual modulation amplitude $\Delta(dR/dE_R)$ of S recoil rates as a function of the recoil energy $E_R$ for $m=100$ GeV$/{\rm c}^2$ and $m=7$ GeV$/{\rm c}^2$, respectively integrated over different hemispheres (colored lines) and over all directions (black line). In the four panels of the figure we use one of  the four different  combinations of $V_{\rm GalRot}=180$ km/s and 312 km/s, and $\sigma_v=173$ km/s and 225 km/s. The escape velocity is always  $v_{\rm esc}=544$ km/s. The red, purple, cyan, and blue curves are for hemispheres having as pole ${\bf V}_{\rm {Earth Rev}}$ on different dates in a year separated by three months. The green and orange curves show the GHAM with largest amplitude at low and high energies, respectively, and they correspond to complimentary hemispheres. For the green lines, the D1$-$D2 dates are Dec 1$-$May 30 in the (a) and (c) panels, and Dec 5$-$June 2 in the (b) and (d) panels. For the orange lines the D1 and D2 dates are exchanged. The annual modulation amplitude (black line) is thus the orange line minus the green line, and goes to zero when they are equal. Notice that the GHAM amplitudes are much larger and become negative at much smaller energies than the usual non-directional annual modulation amplitude.

In Figs.~\ref{Rate-GHAM-m100} and \ref{Rate-GHAM-m7}, one can see that the annual modulation amplitudes strongly depend on $V_{\rm lab}$ and $\sigma_v$, although more work is necessary to quantify this dependence. For smaller $V_{\rm lab}$, the annual modulation amplitude, which is approximately $(V_{\rm EarthRev}/V_{\rm lab})$ times the annual average rate, is larger. A smaller  $\sigma_v$ also leads to a more anisotropic recoil rate, which results in a larger rate difference. Thus the largest modulation amplitudes happen for the smallest $V_{\rm lab}$ and  $\sigma_v$. In Figs.~\ref{Rate-GHAM-m100} and \ref{Rate-GHAM-m7}, the maximum amplitude in the (a) panels are about a factor of two larger than the maximum amplitudes for the largest $V_{\rm lab}$ and  $\sigma_v$, shown in the (d) panels. Not only the magnitude of the amplitudes but their shape as a function of $E_R$ change with $V_{\rm lab}$ and  $\sigma_v$. As the energy increases (and thus $v_q$ increases), for larger $V_{\rm lab}$ or $\sigma_v$, there are more WIMPs with velocity larger than $v_q$, and thus there is a larger fraction of recoil events at higher energies. Therefore, increasing $V_{\rm lab}$ or $\sigma_v$, flattens the recoil spectrum and thus also the rate difference. This can be seen in Figs.~\ref{Rate-GHAM-m100} and \ref{Rate-GHAM-m7} where in panels (a) the slope of the curves beyond the peak is the largest.

Measuring the energy differential rate requires very large statistics. Thus we explore next the GHAM in the energy-integrated rate. Let $R_{\rm max}=\int_{E_1}^{E_2}dE_R (dR_{\rm max}/dE_R)$ and $R_{\rm min}=\int_{E_1}^{E_2}dE_R (dR_{\rm min}/dE_R)$ denote the maximum and minimum energy-integrated rates during a year integrated over a particular energy interval between $E_1$ and $E_2$. The annual modulation amplitude of the energy-integrated rate is then
\begin{equation}
\Delta R=R_{\textrm{max}}-R_{\textrm{min}}=\int_{E_1}^{E_2}{\Delta\left(\frac{dR}{dE_R}\right) dE_R}.
\label{DeltaR}
\end{equation}

\begin{figure}[t]
\begin{center}
  \includegraphics[height=120pt]{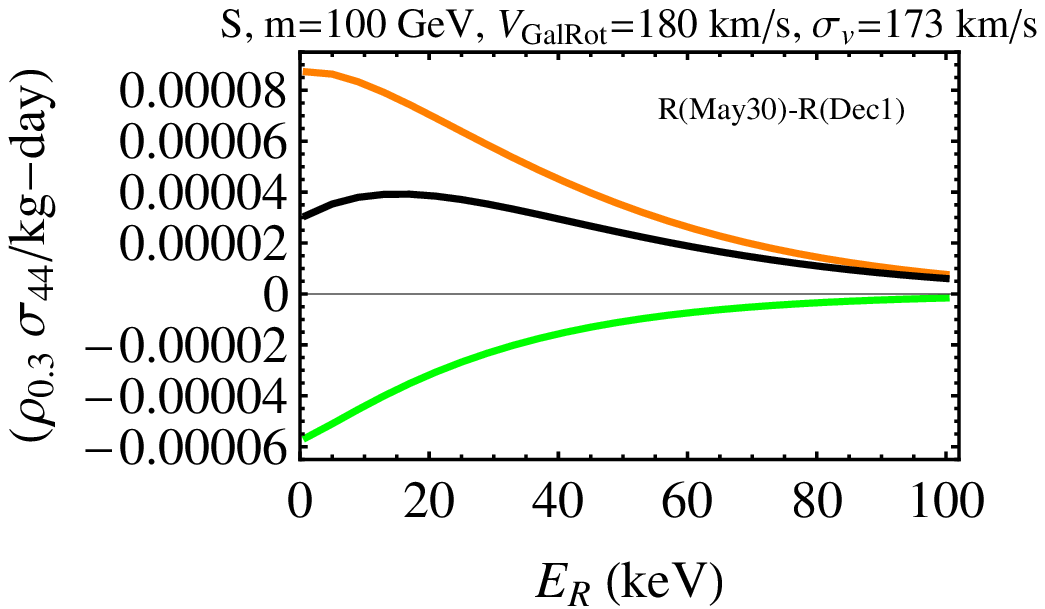}
  \includegraphics[height=120pt]{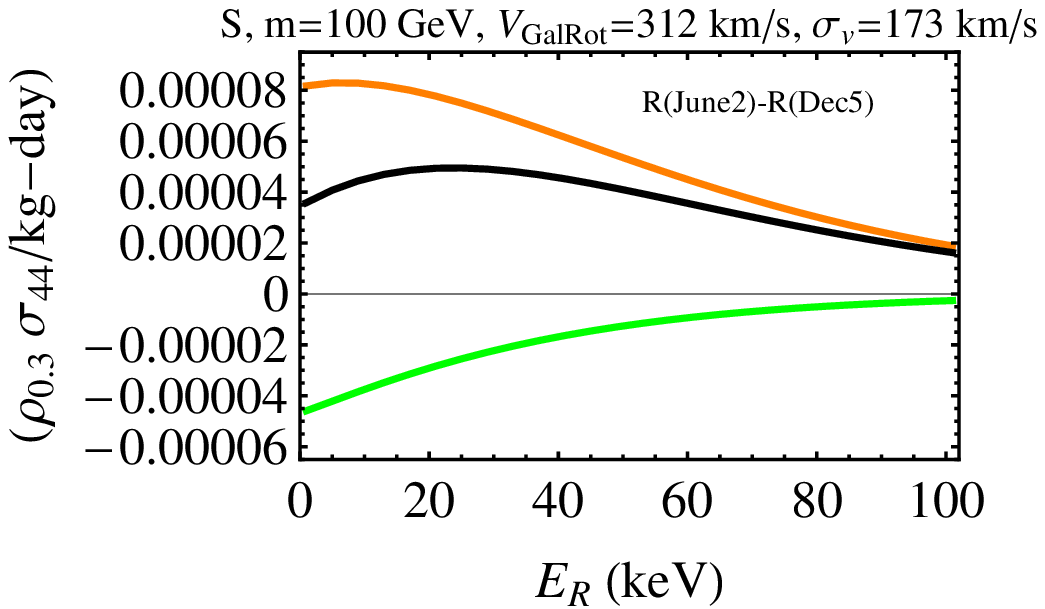}
  \includegraphics[height=120pt]{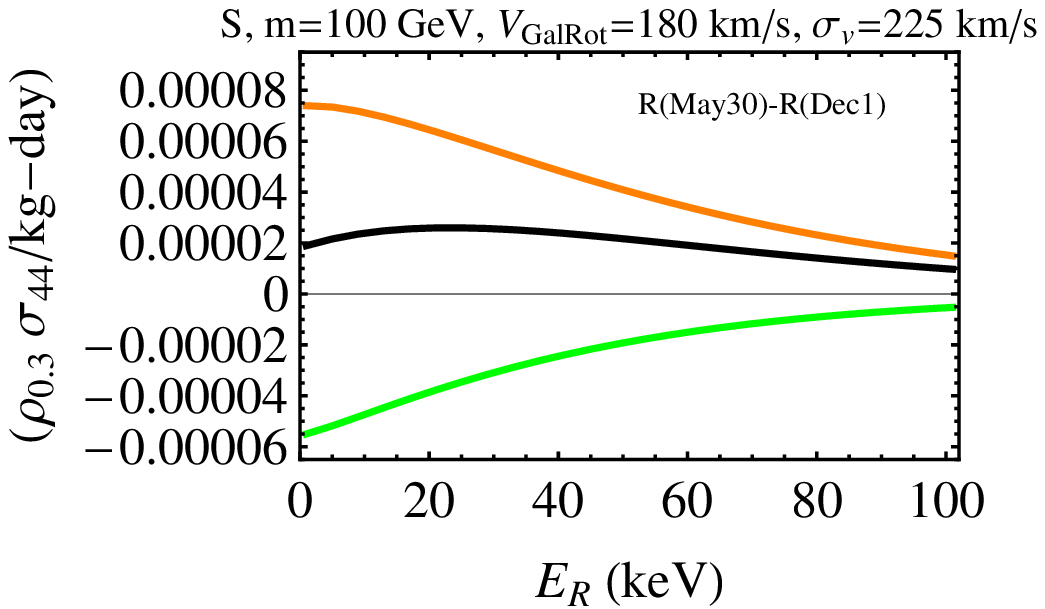}
  \includegraphics[height=120pt]{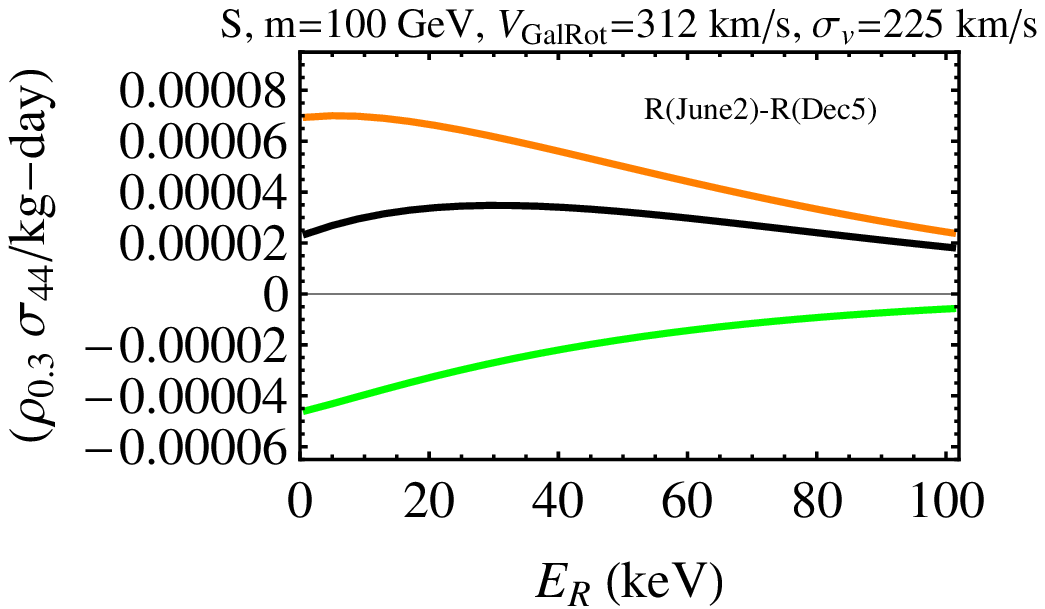}\\
  \vspace{-0.1cm}\caption{Same as Fig.~\ref{Rate-GHAM-m100} but showing the annual modulation amplitude of the energy-integrated recoil rate as a function of the recoil energy $E_R$ and integrated above $E_R$ and over the two hemispheres with the GHAM with largest amplitude at high (orange curve) and low (green curve) energies, as well as over the total sky (black curve).}
  \label{GHAM-EInt}
\end{center}
\end{figure}

Fig.~\ref{GHAM-EInt} shows the annual modulation amplitudes of S recoil rates for $m=100$ GeV$/{\rm c}^2$  integrated above $E_R$, as a function of $E_R$, for the two hemispheres with the GHAM with largest amplitude at high (orange curve) and low (green curve) energies, as well as over the total sky (black curve). As explained before, these two are complimentary hemispheres: in one of them the rate is maximum when $V_{\rm lab}$ is maximum and minimum when $V_{\rm lab}$ is minimum (orange line), and in the other the opposite is true (green line). In Fig.~\ref{GHAM-EInt}, we show the difference between the rates at date D1 and date D2, choosing D1 as the date at which $V_{\rm lab}$ is maximum and D2 as the date at which $V_{\rm lab}$ is minimum. Thus the orange GHAM amplitude is positive, the green is negative, and the usual non-directional annual modulation is the sum of the two. The orange GHAM amplitude reaches values between two and three times larger than the usual non-directional modulation amplitude, at low energies. The parameters used in the four panels are the same as the parameters used in Fig.~\ref{Rate-GHAM-m100}.

In Fig.~\ref{GHAM-EInt} the GHAM with largest amplitude at high energies (orange curve) is larger in magnitude than that at low energies (green curve). This is because the area under the orange curve in Fig.~\ref{Rate-GHAM-m100} (which shows the modulation amplitude of the energy differential recoil rate) is larger than the area under the green curve in the same figure. The dependence of the energy-integrated annual modulation amplitudes on the values of $V_{\rm lab}$ and $\sigma_v$ is similar to that of the energy differential rate (Fig.~\ref{Rate-GHAM-m100}). The peak of the largest GHAM amplitude is largest for the smallest $V_{\rm lab}$ and $\sigma_v$.

\subsection{NGHAM and SGHAM amplitudes}

\begin{figure}[t]
\begin{center}
  \includegraphics[height=130pt]{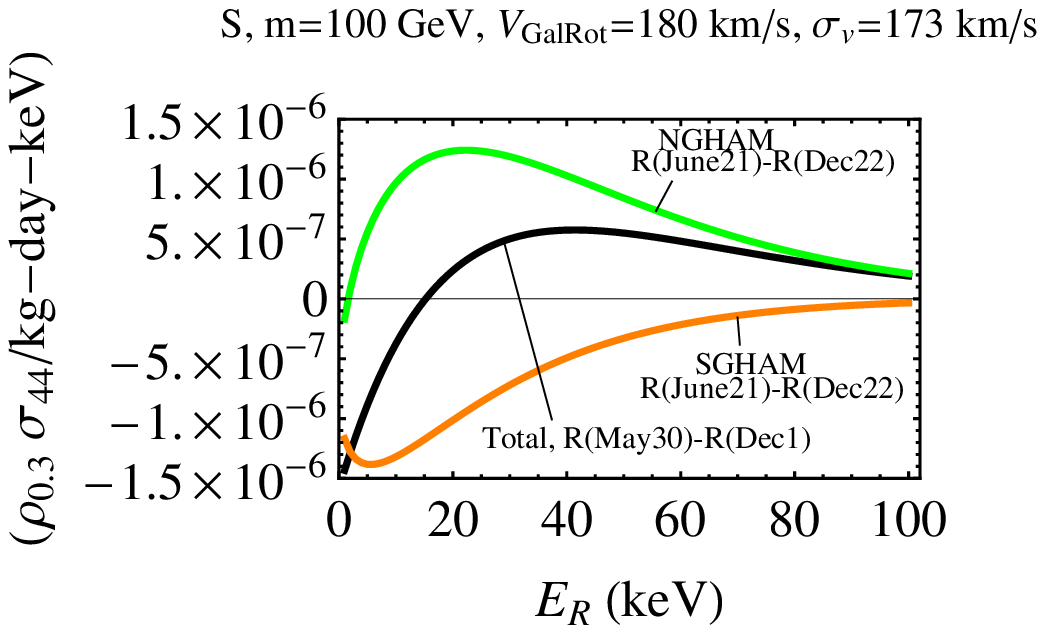}
  \includegraphics[height=130pt]{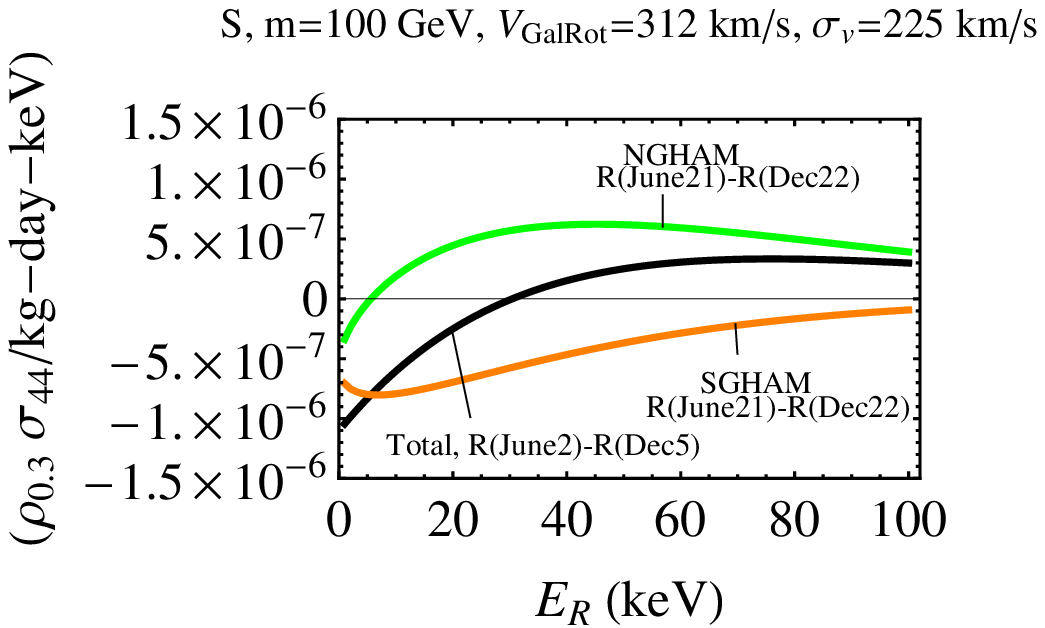}\\
  \vspace{-0.1cm}\caption{Annual modulation amplitude (NGHAM, SGHAM and total) of the S energy differential recoil rate, Eq.~\ref{AnnualMod}, as a function of the recoil energy $E_R$ for $m=100$ GeV/$c^2$, $v_{\rm esc}=544$ km/s and (a) $V_{\rm GalRot}=180$ km/s and  $\sigma_v=173$ km/s (left), and (b) $V_{\rm GalRot}=312$ km/s and $\sigma_v=225$ km/s (right). The black, green and orange curves are for the total (i.e.~non-directional), NGHAM, and SGHAM amplitudes, respectively. Notice that the GHAM amplitudes are much larger than the usual non-directional annual modulation amplitude and change sign at lower energies.}
  \label{Rate-NSGHAM-m100}
\end{center}
\end{figure}

In hemispheres other than those defined in Section 5.1, the GHAM are smaller. As an example, we consider here the North and South Galactic hemispheres, i.e.~the North GHAM  (NGHAM)  and the South GHAM (SGHAM). The  amplitudes  of these modulations are respectively maximum when the component of ${\bf V}_{\rm {EarthRev}}$ towards either the North or South directions (i.e.~in the positive and negative  $z_g$ direction) is maximum, which happens on  December 22  and June 21, respectively. These amplitudes and the total (i.e.~for the whole sky) modulation amplitudes for the S energy differential rate (Eq.~\ref{AnnualMod}) are shown is Fig.~\ref{Rate-NSGHAM-m100} as a function of the recoil energy $E_R$, for $m=100$ GeV$/{\rm c}^2$, $v_{\rm esc}=544$ km/s, and extreme combinations of  $V_{\rm GalRot}$ and  $\sigma_v$, 180 km/s and 173 km/s in the left panel, and $312$ km/s and $225$ km/s in the right. As explained earlier, the combination of small $V_{\rm GalRot}$ and $\sigma_v$ results in the higher modulation amplitudes, whereas the opposite happens for the combination of large $V_{\rm GalRot}$ and $\sigma_v$.

We find that both GHAM amplitudes in Fig.~\ref{Rate-NSGHAM-m100} are smaller than the GHAM amplitudes of the previous section (Fig.~\ref{Rate-GHAM-m100}), but still larger and change sign at much lower energies than the total sky amplitude. The NGHAM amplitude is positive except at very low energies and the SGHAM amplitude  is negative, having chosen to present the difference in rates at the same two dates. This is what one expects because the mean recoil direction, the direction of $-{\bf V}_{\rm lab}$, points slightly North in June and slightly South in December (see Figs.~\ref{Vlab-plot}, \ref{Flux} and \ref{m100e5-180-173}). Thus going from June to December, the recoil rate decreases in the NGH and increases in the SGH.

\section{Statistical test of detectability of the GHAM}

Here we estimate the minimum number of events necessary to detect the GHAM. Let $N_h$ and $N_l$ denote the number of recoil events during the annual half-cycles with high, and low rates respectively, in a particular recoil energy interval between $E_1$ and $E_2$. The number of events in both cycles is $N=N_h+N_l$. For the annual modulation to be observable at the $3\sigma$ level we require
\begin{equation}
N_h - N_l > 3 \sigma \simeq 3 \sqrt{N/2},
\label{3sigma-annual}
\end{equation}
where $\sigma \simeq \sqrt{N/2}$ is the standard deviation of $N_h$ or $N_l$, given that $N_h \simeq N_l$ for a small annual modulation. The energy-integrated rate can be written as $R(t) = R_0 + R_m \cos (\omega t)$ with $\omega=2\pi/$(1 yr). The unmodulated part is $R_0=(R_{\rm max}+R_{\rm min})/2$ and the modulated part has amplitude $R_m=(R_{\rm max}-R_{\rm min})/2$. If the detector exposure is $M T$, then  $N_h = MT (R_0/2 + R_m/\pi)$ and  $N_l = M T (R_0/2 - R_m/\pi)$. The factors of $1/\pi$ come from integrating the modulated part of the rate, $R_m \cos (\omega t)$ over a half-cycle.

We define the relative modulation amplitude $A$ (taking into account the signal only and neglecting background) in terms of the maximum and minimum recoil rates integrated over energy and direction,
 \begin{equation}
 A= \frac{R_{\rm max}-R_{\rm min}}{R_{\rm max}+R_{\rm min}}= \frac{R_m}{R_0}.
\label{A}
 \end{equation}
Thus, Eq.~\ref{3sigma-annual} becomes
\begin{equation}
N > \frac{9 \kappa}{A^2} = N_{\rm min},
\label{N-Cond}
\end{equation}
where $\kappa = \pi^2/8$ for a sinusoidal modulation (it would be slightly different otherwise). Alternatively, Eq.~\ref{N-Cond} can be written as
\begin{equation}
MT > \frac{9 \kappa}{A^2 R_0}=\frac{18 \kappa}{A^2 (R_{\rm max}+R_{\rm min})} =(MT)_{\rm min}.
\label{MT-Cond}
\end{equation}

We can apply this condition to any annual modulation. The condition for the observability of the GHAM is $N_H > 9 \kappa/A_H^2$, where $N_H$ is the number of events towards one particular Galactic hemisphere H, and $A_H$ is the relative modulation amplitude for that Galactic hemisphere. One can also find the particular exposure $(MT)_{H, {\rm min}}= 9 \kappa/(A_H^2 R_{0,H})$ for the hemisphere H, where $R_{0,H}$ is the average rate in the hemisphere H. Using the average rate $R_{0,{H'}}$ in the complimentary hemisphere H$'$, one can find the minimum number of events needed with $(MT)_{H, {\rm min}}$ in the complimentary hemisphere H$'$, $N_{H',{\rm min}}=N_{H,{\rm min}}R_{0,{H'}}/R_{0,H}$. Thus the minimum number of events observed in all directions necessary to detect the GHAM in the initial hemisphere H is
\begin{equation}
N_{\rm min}^{\rm GHAM}=N_{H,{\rm min}}+N_{H',{\rm min}}.
\label{N-Cond-tot}
\end{equation}

  \begin{figure}[t]
\begin{center}
  \includegraphics[height=140pt]{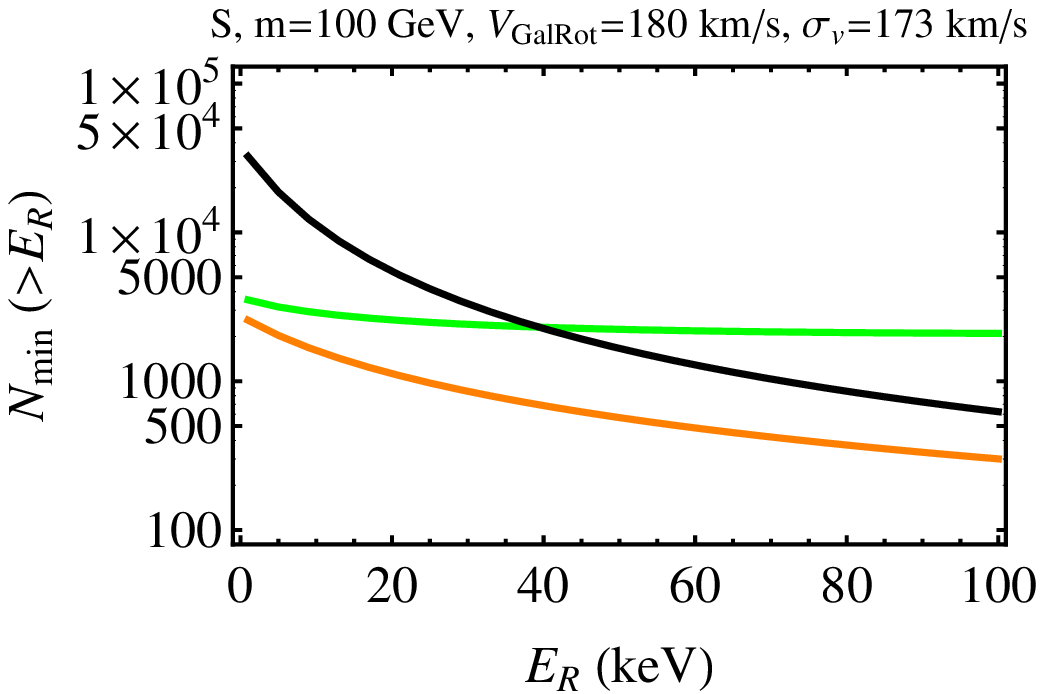}
  \includegraphics[height=140pt]{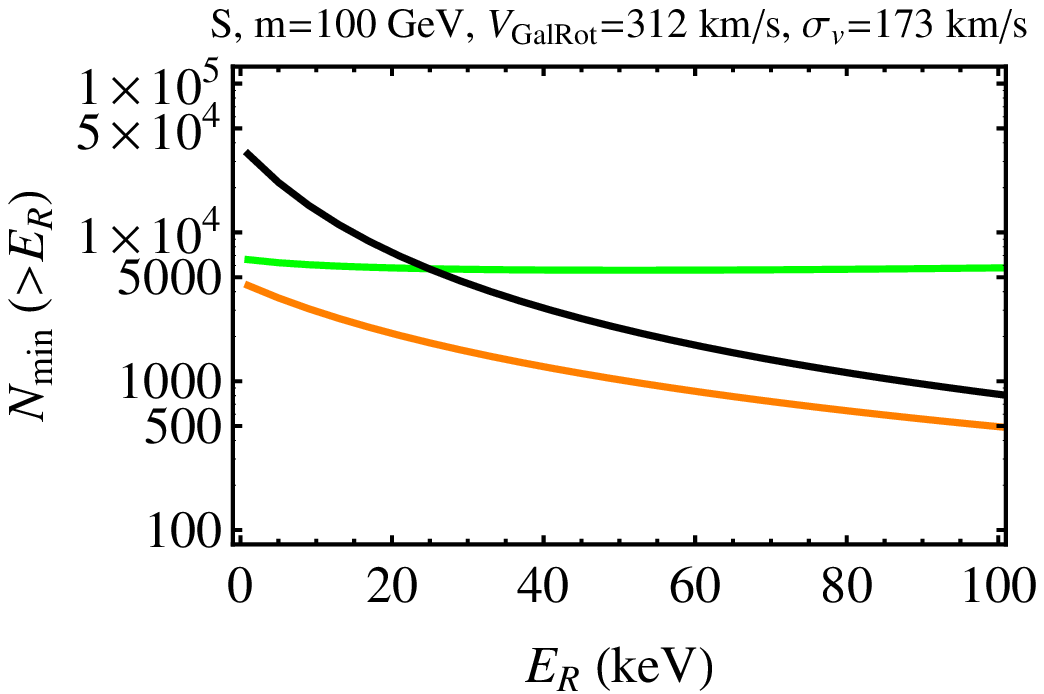}
  \includegraphics[height=140pt]{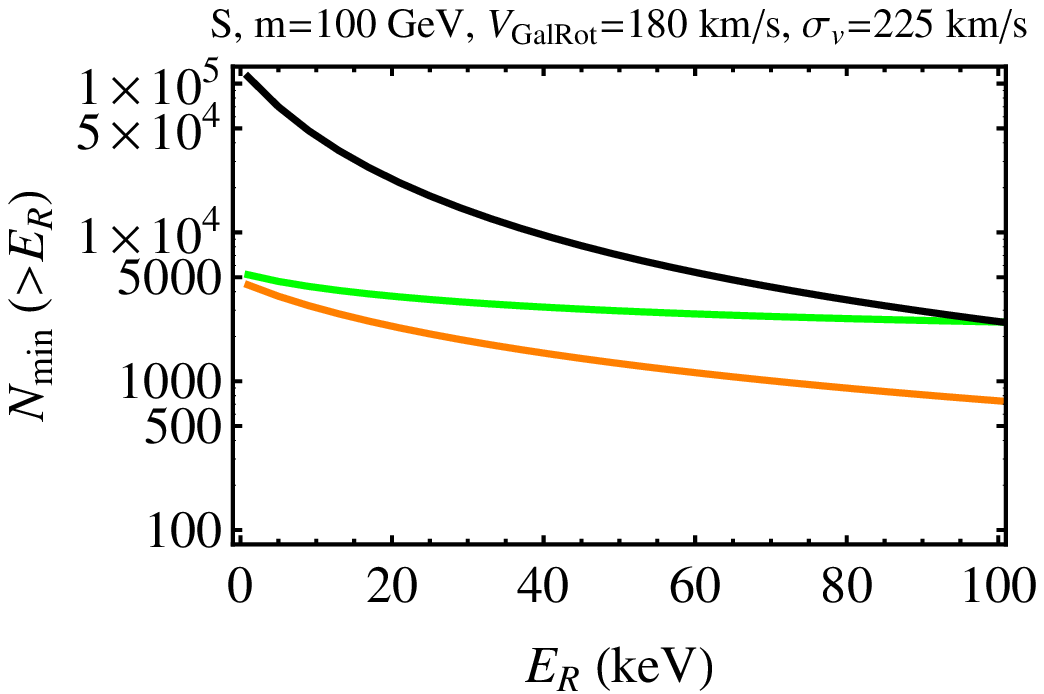}
  \includegraphics[height=140pt]{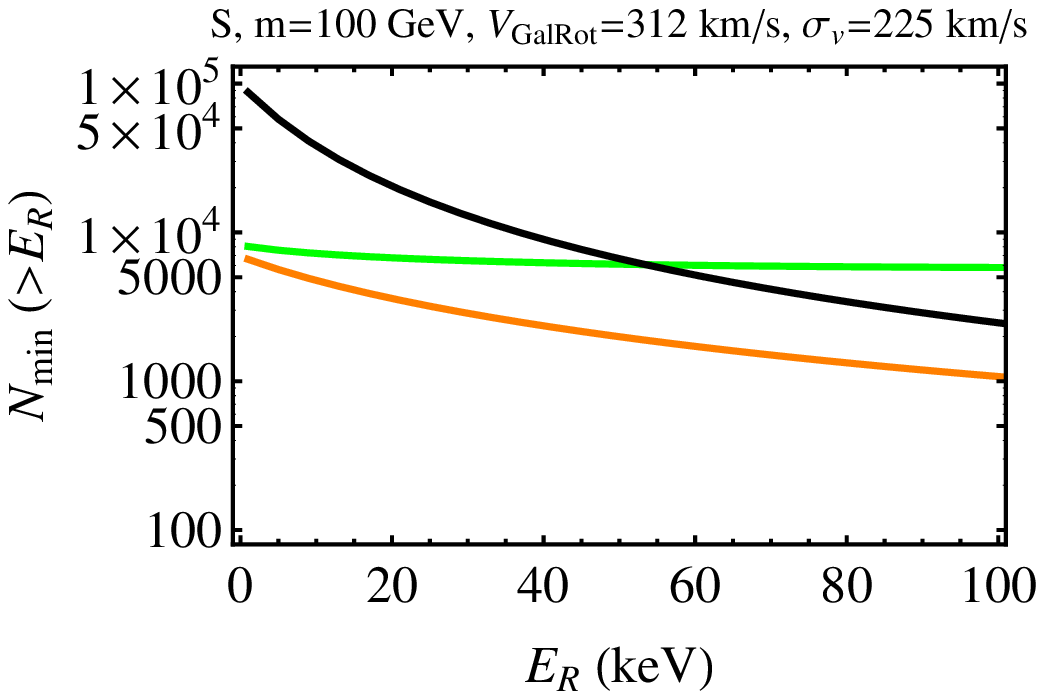}\\
  \vspace{-0.1cm}\caption{Minimum number of events with energy larger than $E_R$ (Eqs.~\ref{N-Cond} and \ref{N-Cond-tot}) in all directions needed to detect the usual non-directional annual modulation (black curve) and the GHAM with largest amplitude at low (green curve) and high (orange curve) energies in an S detector. The parameters used are $m=100$ GeV$/{\rm c}^2$, $v_{\rm esc}=544$ km/s and (a) $V_{\rm GalRot}=180$ km/s and  $\sigma_v=173$ km/s (top left), (b) $V_{\rm GalRot}=312$ km/s and $\sigma_v=173$ km/s (top right), (c) $V_{\rm GalRot}=180$ km/s and $\sigma_v=225$ km/s (bottom left), and (d) $V_{\rm GalRot}=312$ km/s and $\sigma_v=225$ km/s (bottom right).}
  \label{Nmin}
\end{center}
\end{figure}
  \begin{figure}
\begin{center}
  \includegraphics[height=140pt]{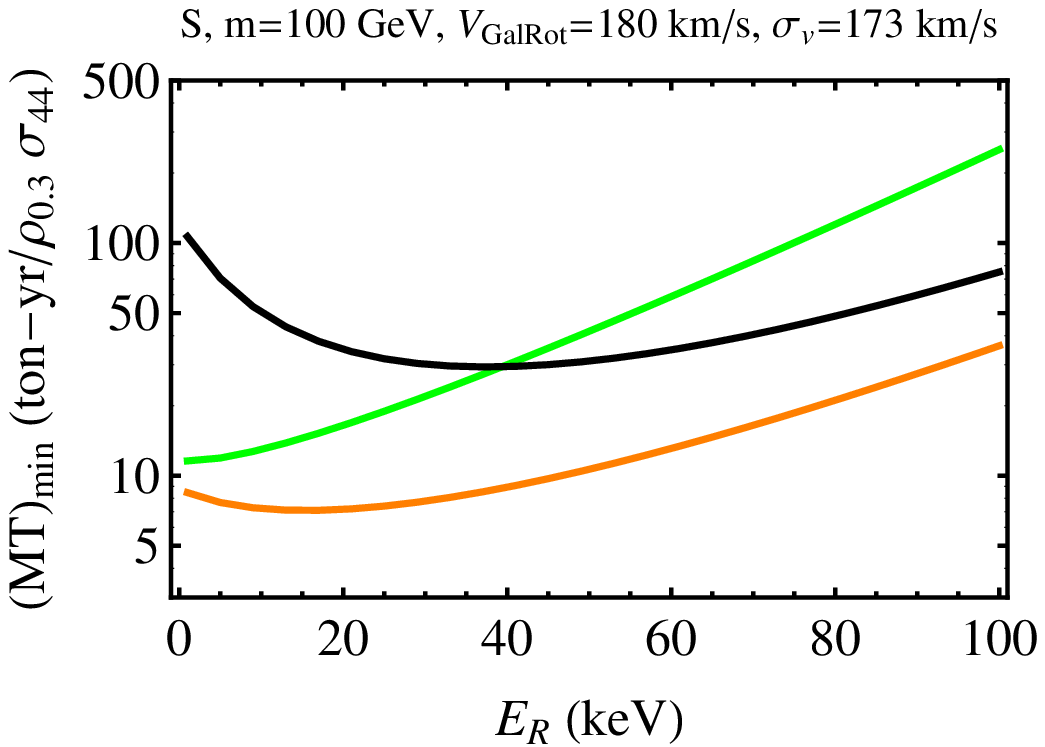}
  \includegraphics[height=140pt]{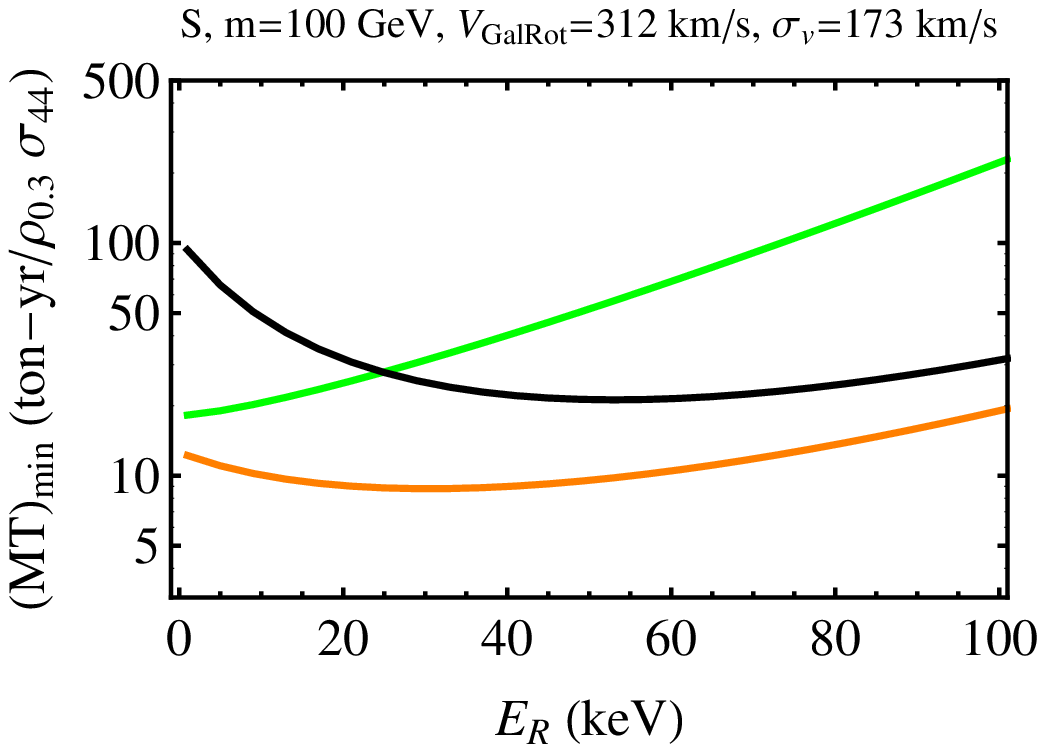}
  \includegraphics[height=140pt]{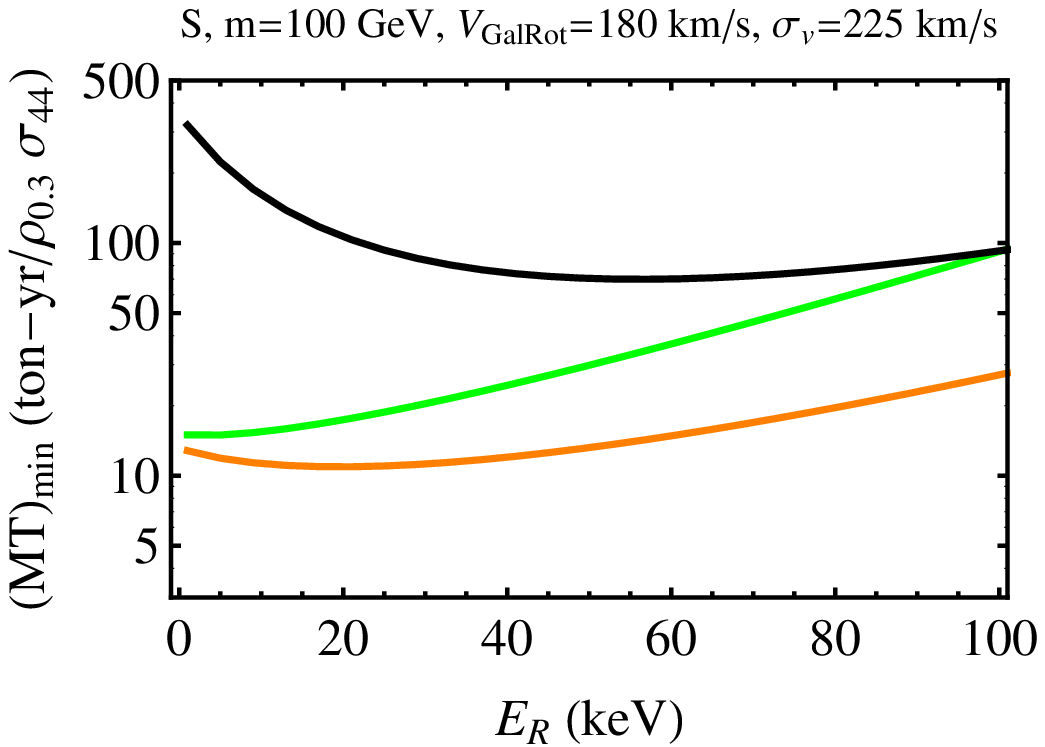}
  \includegraphics[height=140pt]{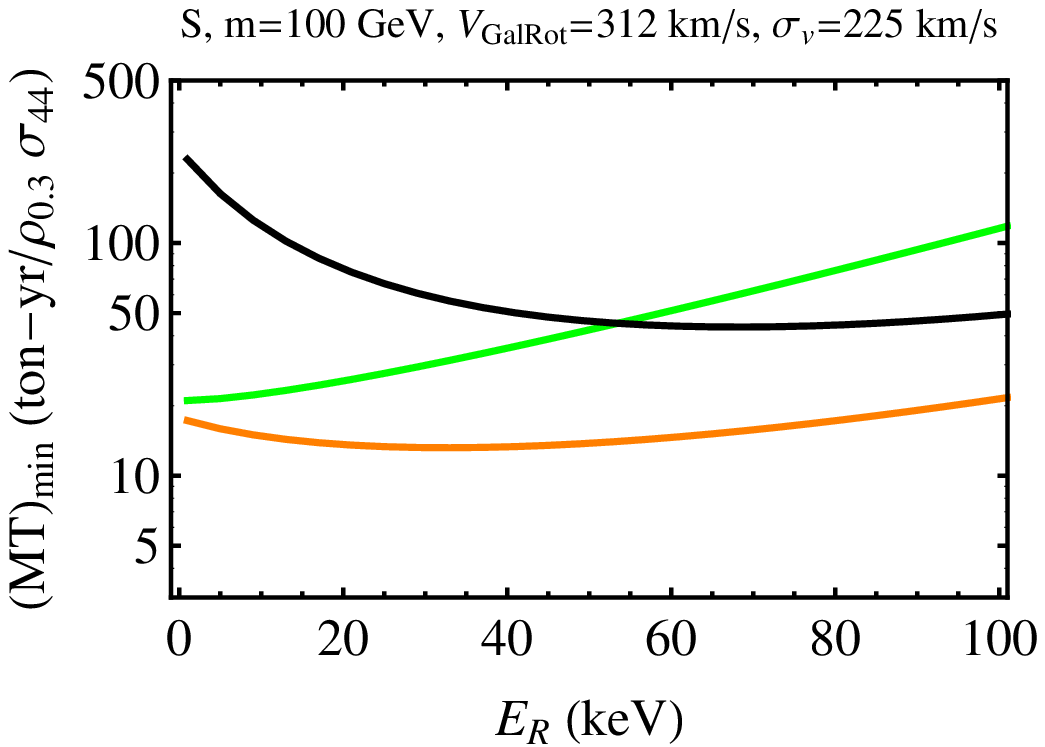}\\
  \vspace{-0.1cm}\caption{Exposure $MT$ (in ton-yr) corresponding to the minimum number of events of Fig.~\ref{Nmin} needed to detect the usual non-directional annual modulation (black curve) and the GHAM with largest amplitude at low (green curve) and high (orange curve) energies, assuming $\rho=0.3$ GeV/$c^2$/cm$^3$ and $\sigma_p=10^{-44}\;{\text{cm}}^2$.}
  \label{MTmin}
\end{center}
\end{figure}
  \begin{figure}
\begin{center}
  \includegraphics[height=135pt]{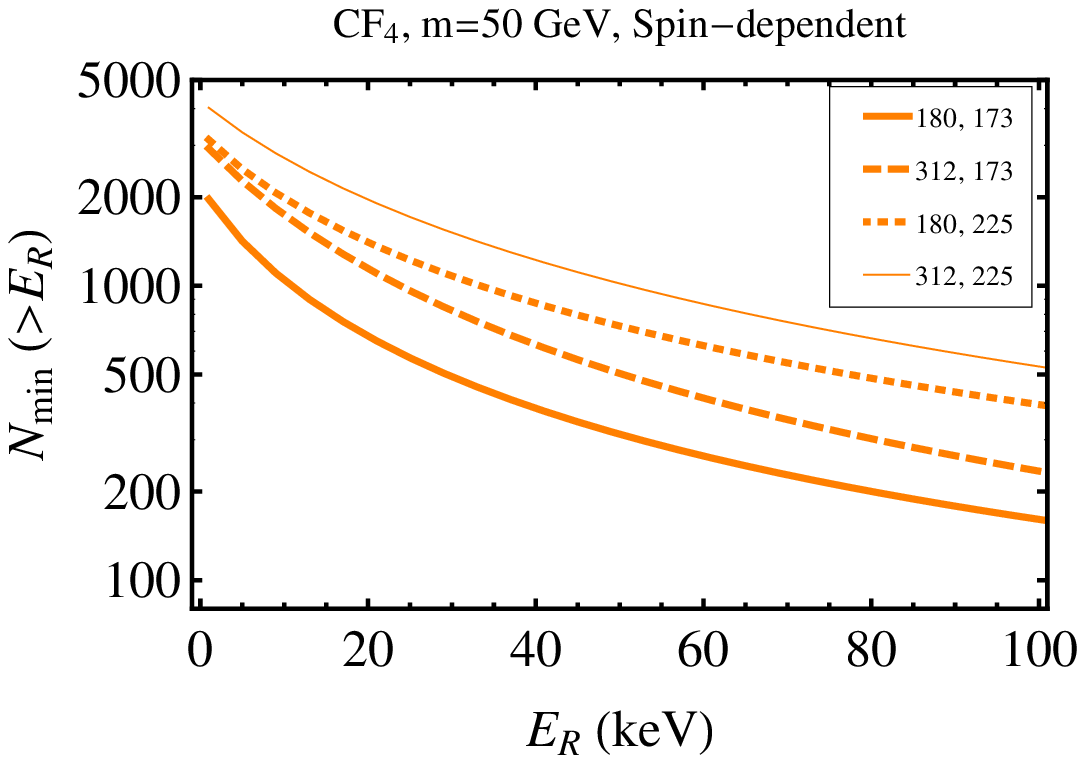}
  \includegraphics[height=135pt]{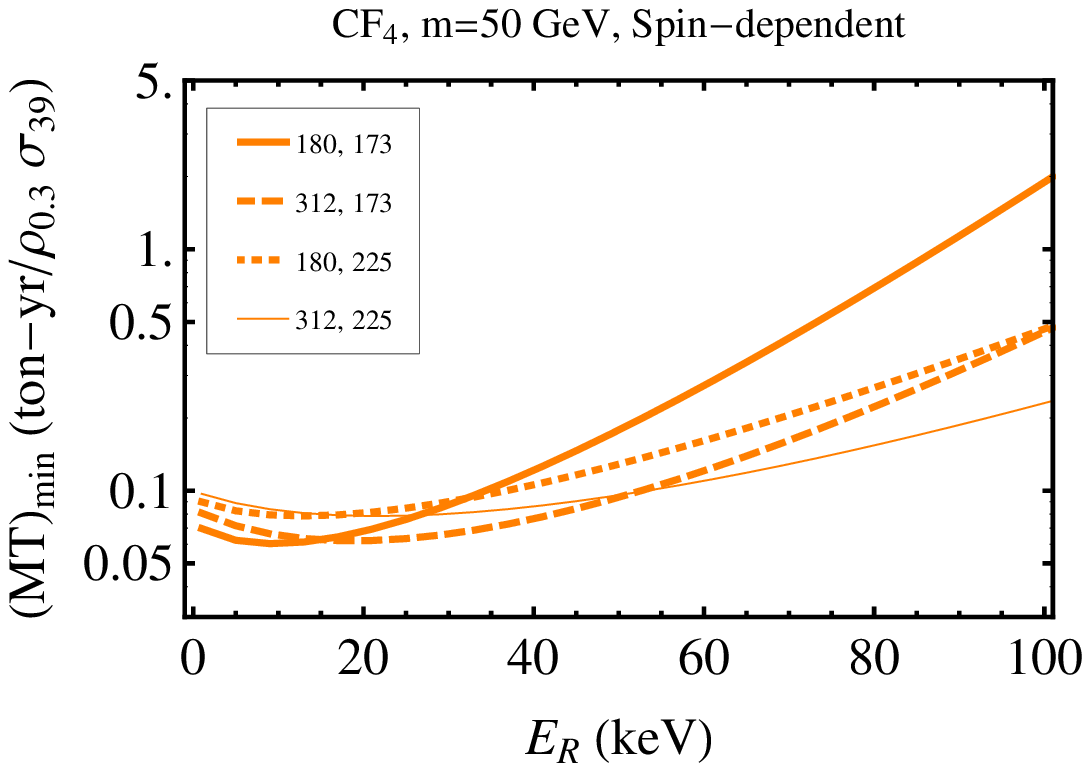}\\
  \vspace{-0.1cm}\caption{(a) (Left) Minimum number of events (in all directions) and (b) (right) minimum exposure $MT$ (in ton-yr) needed to detect the GHAM with largest amplitude at high energies in the SD rate integrated above $E_R$ in CF$_4$. The different curves in each plot are for different values of $V_{\rm GalRot}$ and $\sigma_v$, 180 km/s and 173 km/s (thick solid), 312 km/s and 173 km/s (dashed), 180 km/s and 225 km/s (dotted), and 312 km/s and 225 km/s (thin solid), respectively. In (b) we assume $\rho=0.3$ GeV/$c^2$/cm$^3$ and $\sigma_p^{\rm SD}=10^{-39}\;{\text{cm}}^2$.}
  \label{N-MTmin-SD}
\end{center}
\end{figure}

To see if the GHAM would be easier to detect compared to the total annual modulation, we need to compare the minimum number of events $N_{\rm min}$ necessary to detect the total annual modulation and $N_{\rm min}^{\rm GHAM}$ in all directions necessary to detect the GHAM.

Fig.~\ref{Nmin} shows the minimum number of S recoils detected (in all directions) with energy larger than $E_R$, Eqs.~\ref{N-Cond} and \ref{N-Cond-tot}, needed to detect the total annual modulation (black curve), and the GHAM with the largest amplitude at low (green curve) and high (orange curve) energies. In the four panels of the figure we use one of  the four different  combinations of $V_{\rm GalRot}=180$ km/s and 312 km/s, and $\sigma_v=173$ km/s and 225 km/s, for $m=100$ GeV$/{\rm c}^2$ and $v_{\rm esc}=544$ km/s.
Fig.~\ref{Nmin} shows that one needs a minimum of a few thousand events to observe the largest GHAM amplitudes, and one would need at least 10 times more to observe the usual non-directional annual modulation.

Notice that although $N_{\rm min}$ is smaller at larger energies, the average rate $R_0$ decreases with energy, thus the necessary exposure increases at higher energies. This can be seen in Fig.~\ref{MTmin} which shows the minimum exposure (Eq.~\ref{MT-Cond}) as function of $E_R$ needed to detect the non-directional annual modulation (black curve), and the GHAM with largest amplitude at low (green curve) and high (orange curve) energies in the rate integrated above $E_R$ for an S detector. The parameters used are the same as those in Fig.~\ref{Nmin}.
Because $A^2$ increases and $R_0$ decreases with energy, the increase or decrease of $(MT)_{\rm min}$ with energy depends on the rate of change with energy of $A^2$ and $R_0$. Thus $(MT)_{\rm min}$ has a minimum at a certain recoil energy.

Fig.~\ref{N-MTmin-SD}.a shows the minimum number of events in all directions, with energy above $E_R$, needed to detect the GHAM with largest amplitude at high energies in a CF$_4$ detector for a 50 GeV$/{\rm c}^2$ WIMP with SD interactions. We can use SD cross sections larger than the SI cross sections ($\sigma_p^{\rm SD}=10^{-39}\;{\text{cm}}^2$) without violating experimental bounds, which results in smaller exposures needed to detect the GHAM. Fig.~\ref{N-MTmin-SD}.b shows the minimum exposure needed  to detect the GHAM with largest amplitude at high energies in the SD rate integrated above $E_R$ in CF$_4$. The four curves in Fig.~\ref{N-MTmin-SD} are for the four different combinations of $V_{\rm GalRot}=180$ km/s and 312 km/s, and $\sigma_v=173$ km/s and 225 km/s. Fig.~\ref{N-MTmin-SD}.a shows that one needs between a few hundred and a few thousand events depending on the value of $V_{\rm GalRot}$ and $\sigma_v$ and the energy interval, to observe the largest GHAM amplitude. From Fig.~\ref{N-MTmin-SD}.b one can see that the minimum exposure needed to detect the largest GHAM in the SD rate is between 0.06 ton-yr and a few ton-yr. Figs.~\ref{Nmin} and \ref{N-MTmin-SD}.a show that the combination of low $V_{\rm GalRot}$ and $\sigma_v$ gives the smallest minimum number of events.

\section{Anisotropic Logarithmic-ellipsoidal models}

\begin{figure}
\begin{center}
  \includegraphics[height=130pt]{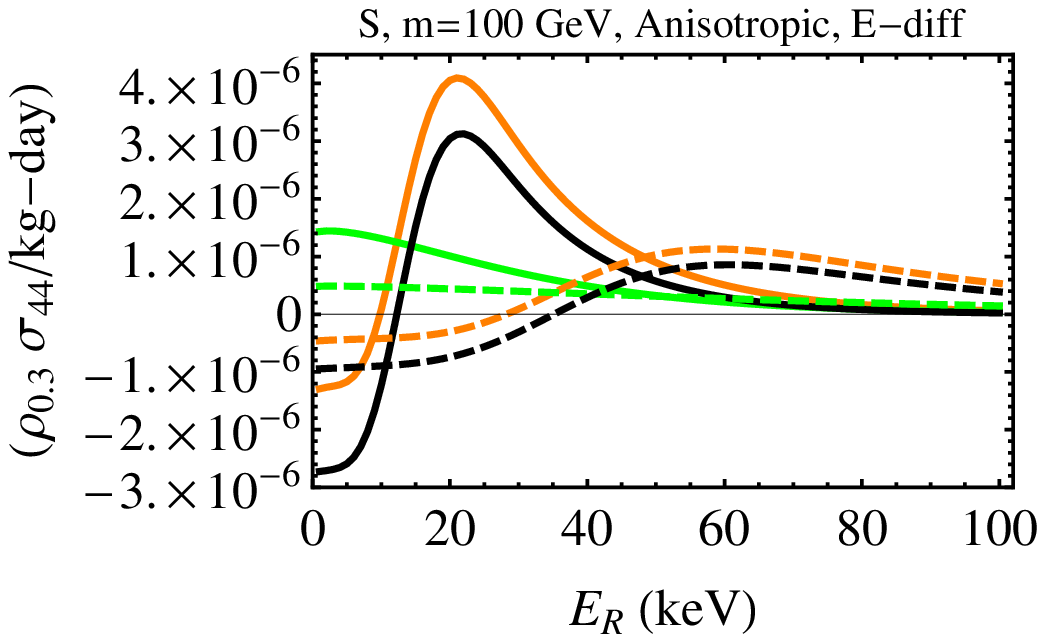}
  \includegraphics[height=130pt]{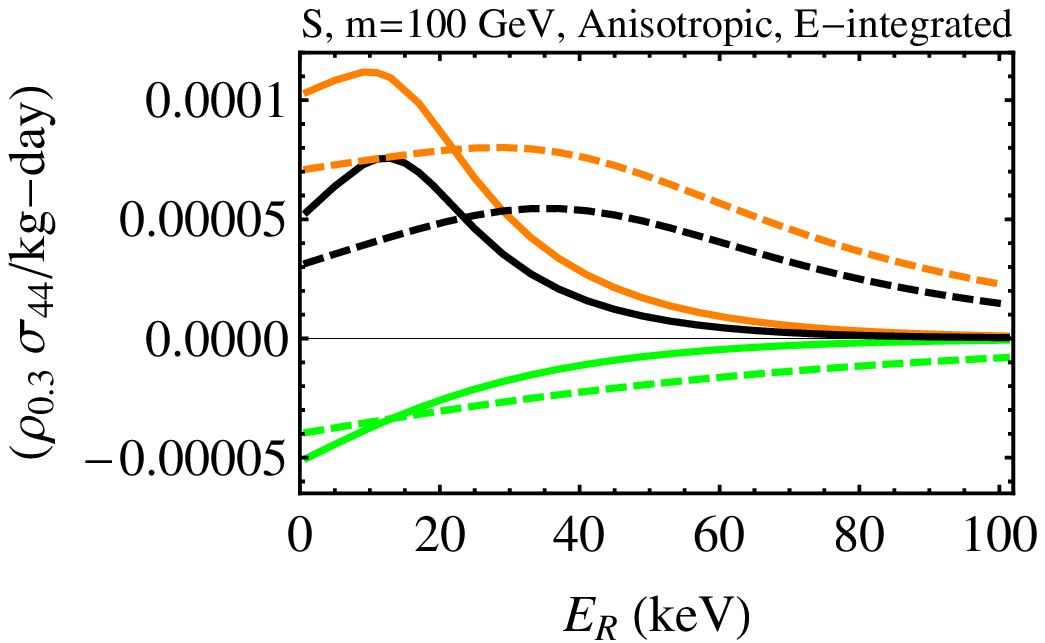}
  \includegraphics[height=140pt]{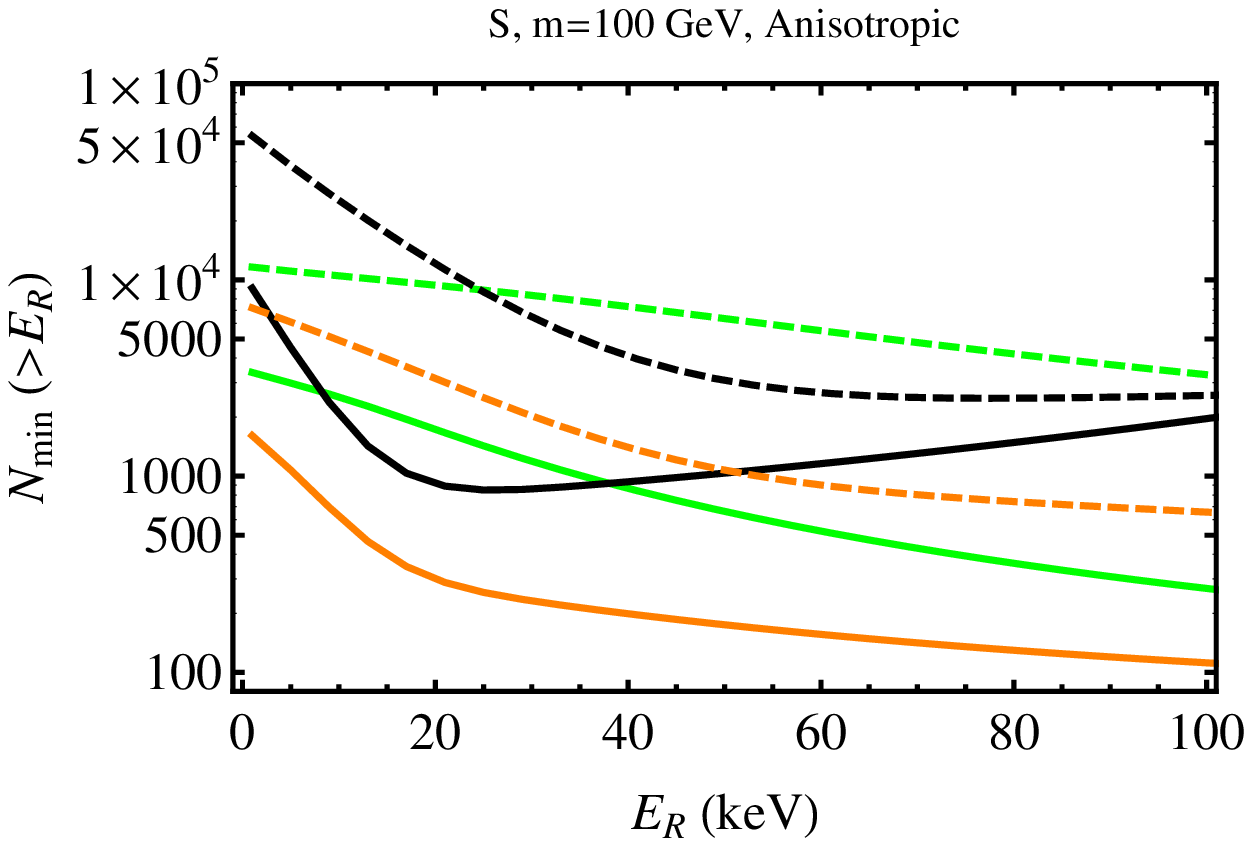}
  \includegraphics[height=145pt]{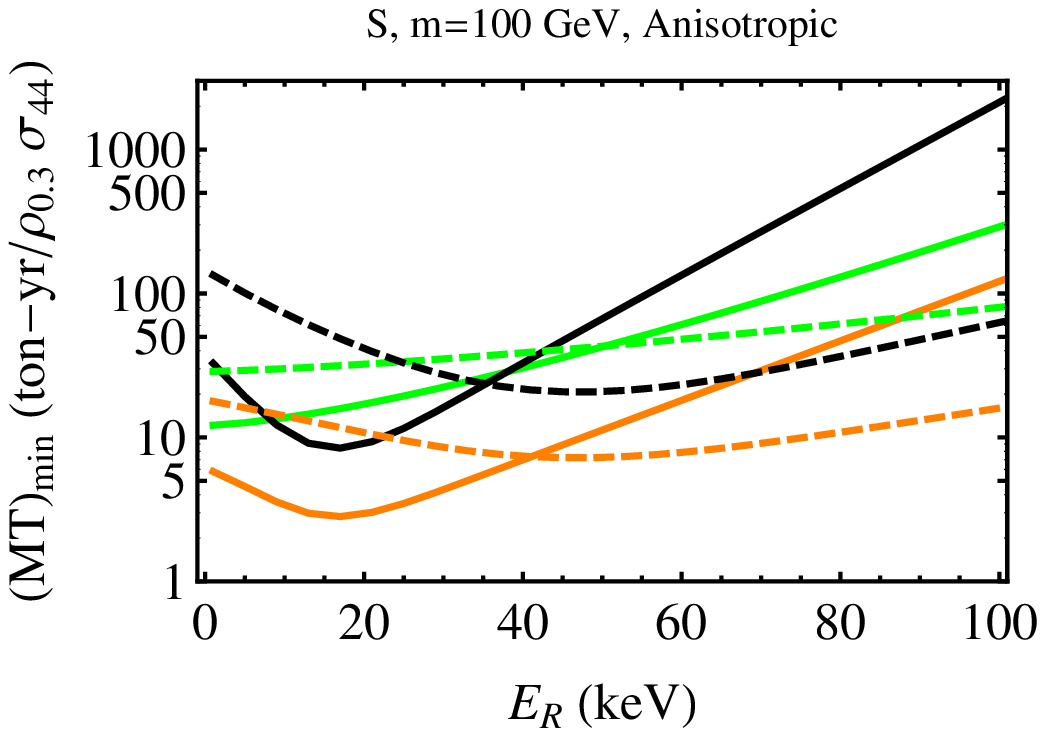}\\
  \vspace{-0.1cm}\caption{An example of an anisotropic logarithmic-ellipsoidal model with $\mathfrak{p}=0.72$, $\mathfrak{q}=0.7$, $\Gamma=4.02$, and $m=100$ GeV$/{\rm c}^2$, $V_{\rm GalRot}=180$ km/s (solid curves), and $V_{\rm GalRot}=312$ km/s (dashed curves). (a) (Top left) Annual modulation amplitude of the S energy differential rate integrated over the two hemispheres with the GHAM with largest amplitude at high (orange curves) and low (green curves) energies, as well as over the total sky (black curves). For the solid and dashed green curves, the D1$-$D2 dates are Dec 1$-$May 30 and Dec 5$-$June 2, respectively. For the orange curves the D1 and D2 dates are exchanged. (b) (Top right) Same as (a) but showing the annual modulation amplitude of the energy-integrated rate as a function of the recoil energy $E_R$ and integrated above $E_R$ and over the same two hemispheres, as well as over the total sky. The solid curves are for May 30$-$Dec 1, and the dashed curves are for June 2$-$Dec 5. (c) (Bottom left) Minimum number of events (in all directions) and (d) (bottom right) minimum exposure $MT$ (in ton-yr) needed to detect the total annual modulation (black curves), and the GHAM with the largest amplitude at low (green curves) and high (orange curves) energies in the rate integrated above $E_R$ in S. In (d) we assume $\rho=0.3$ GeV/$c^2$/cm$^3$ and $\sigma_p=10^{-44}\;{\text{cm}}^2$.}
  \label{GHAM-N-MT-Anisot}
\end{center}
\end{figure}

So far we have used the isotropic Maxwell-Boltzmann model. While the details of the velocity distribution are important for the detectability of aberration effects, their existence depends only on the motion of the Earth around the Sun, and thus the aberration effects are general for any WIMP distribution. A complete analysis of aberration effects in non-Maxwellian or non-isotropic distributions is beyond the scope of this paper. However, in Fig.~\ref{GHAM-N-MT-Anisot} we present one such example: an anisotropic logarithmic-ellipsoidal model of Ref.~\cite{Evans:2000} (discussed in Section 5 and Appendix B of Ref.~\cite{Ring-like}). The equations for the velocity dispersions in the direction of the Galactic center, the Galactic rotation, and the north Galactic pole, $\sigma_{xg}$,  $\sigma_{yg}$, and $\sigma_{zg}$, as well as the constants $\mathfrak{p}$ and $\mathfrak{q}$ used to describe the axis ratios of the density ellipsoid, and the constant $\Gamma$ used to describe the velocity anisotropy, are given in Appendix B of Ref.~\cite{Ring-like}. In the example presented in Fig.~\ref{GHAM-N-MT-Anisot}, the Solar system is on the minor axis with $\mathfrak{p}=0.72$, $\mathfrak{q}=0.7$, and $\Gamma=4.02$. For $V_{\rm GalRot}=180$ km/s, this choice of parameters corresponds to $\sigma_{xg}=134$ km/s,  $\sigma_{yg}=32$ km/s, and $\sigma_{zg}=177$ km/s, and for $V_{\rm GalRot}=312$ km/s, it corresponds to $\sigma_{xg}=233$ km/s,  $\sigma_{yg}=56$ km/s, and $\sigma_{zg}=307$ km/s. In Fig.~\ref{GHAM-N-MT-Anisot} $m=100$ GeV$/{\rm c}^2$, and the solid and dashed curves are for $V_{\rm GalRot}$, 180 km/s and 312 km/s, respectively.

Fig.~\ref{GHAM-N-MT-Anisot}.a shows the annual modulation amplitude $\Delta(dR/dE_R)$ of S recoil rates as a function of $E_R$ integrated over the two hemispheres with the GHAM with largest amplitude at high (orange curve) and low (green curve) energies, as well as over the total sky (black curve). For the solid and dashed green curves, the D1$-$D2 dates are Dec 1$-$May 30 and Dec 5$-$June 2, respectively. For the orange curves the D1 and D2 dates are exchanged. Fig.~\ref{GHAM-N-MT-Anisot}.b shows the annual modulation amplitudes integrated above $E_R$, as a function of $E_R$, for the two hemispheres mentioned and over the total sky. In this figure, the solid curves are for May 30$-$Dec 1, and the dashed curves are for June 2$-$Dec 5. Figs.~\ref{GHAM-N-MT-Anisot}.c and \ref{GHAM-N-MT-Anisot}.d show the minimum number of S recoils detected in all directions, and the minimum exposure, respectively, needed to detect the total annual modulation (black curves), and the GHAM with the largest amplitude at low (green curves) and high (orange curves) energies in the rate integrated above $E_R$.

One can see that the magnitude and shape of the annual modulation amplitudes as a function of $E_R$ in the anisotropic model assumed (Figs.~\ref{GHAM-N-MT-Anisot}.a and \ref{GHAM-N-MT-Anisot}.b) are quite different from those in the IMB (Figs.~\ref{Rate-GHAM-m100} and \ref{GHAM-EInt}). In this anisotropic logarithmic-ellipsoidal model, the velocity dispersions depend on $V_{\rm GalRot}$, and thus the GHAM amplitudes strongly depend on $V_{\rm lab}$. Similar to the IMB, the largest modulation amplitudes happen for the smallest $V_{\rm lab}$, and increasing $V_{\rm lab}$, flattens the recoil spectrum and thus also the rate difference. Fig.~\ref{GHAM-N-MT-Anisot}.c and \ref{GHAM-N-MT-Anisot}.d show that one needs a minimum of a few hundred events corresponding to at least a few ton-yr of S to observe the largest GHAM amplitudes in this anisotropic model. This number of events and the corresponding exposure are of the same order of magnitude as for the isotropic Maxwellian case.

\section{Summary}

In this paper we study changes in the recoil direction pattern in directional detectors, which we call aberration after a similar effect on the position of stars in the sky. These aberration features in a dark matter signal are due to the change in the direction of the arrival velocity of WIMPs on Earth, caused by Earth's motion around the Sun. These features depend on the orbital characteristics of the Earth's revolution and on the WIMP velocity distribution. Knowing the former, aberration features can be used to determine the latter, besides being a curious way of observing Earth's revolution and confirming the galactic provenance of WIMPs. While observing the full aberration pattern would require extremely large exposures, we show that the annual variation of the mean recoil direction or of the event counts over specific solid angles may be detectable with moderately large exposures.

The maximal angular annual  separation $\gamma_s$ between the mean recoil directions in a year is inversely proportional to the magnitude of the average WIMP velocity with respect to Earth, $-\bf{V}_{\rm lab}$. The largest uncertainty in $\bf{V}_{\rm lab}$ stems from the Galactic rotation speed at the position of the Sun, $V_{\rm GalRot}$. Thus measuring $\gamma_s$ would allow to determine $V_{\rm GalRot}$. The dependence of $\gamma_s$ on $V_{\rm GalRot}$ is shown in Fig.~\ref{AngularSep}. Depending on the value of $V_{\rm {GalRot}}$, between 180 km/s and 312 km/s, $\gamma_s$ could be between 18$^\circ$ and 11$^\circ$. In Section 4, we estimated that between 1400 and a few thousand events would be needed to measure the angular difference in recoil direction integrated over six or three months. We used as the basis for our estimate, the minimum number of events given in Refs.~\cite{Lee:2012pf} and \cite{Billard:2010jh} needed to determine the mean recoil direction with an error of a few degrees. The error in the determination of $V_{\rm {GalRot}}$ due to an uncertainty in $\gamma_s$ can be read off of Fig.~\ref{AngularSep}. For example, with an error of $3.5^\circ$, the measured $V_{\rm GalRot}$ would be $180^{+60}_{-30}$ km/s or $312^{+190}_{-70}$ km/s using the three month averaged curve, and $180^{+90}_{-40}$ km/s or $312^{+290}_{-110}$ km/s using the six month averaged curve.

In Section 5 we showed that with more than a thousand events, it would also be possible to detect the annual modulation due to aberration of the energy integrated rate. The integrated counts over Galactic hemispheres separated by planes perpendicular to Earth's orbit would modulate annually, resulting in Galactic Hemisphere Annual Modulations (GHAM), with amplitudes larger than the usual non-directional annual modulation. The Galactic hemispheres with the largest GHAM are those divided by planes perpendicular to the Earth's orbital velocity when $V_{\rm lab}$ is maximum (see Figs.~\ref{Rate-GHAM-m100}, \ref{Rate-GHAM-m7}). In this case the full orbital velocity is in the direction of one hemisphere at one time and away from it six months later. In the examples shown in Figs.~\ref{Rate-GHAM-m100}, \ref{Rate-GHAM-m7} and \ref{GHAM-EInt}, the amplitudes of the GHAM are larger than those of the usual annual modulation of the non-directional rate by factor of at least two, and for the energy differential rate change sign at much lower energies than the usual non-directional annual modulation amplitude. As shown in Fig.~\ref{MTmin}, assuming no background and perfect energy and angular  experimental resolution, at least a few ton-yr of S would be required to detect the GHAM of the spin-independent energy-integrated rate. To detect the largest GHAM of the spin-dependent energy-integrated rate in CF$_4$, a minimum exposure of 60 kg-yr would be required as shown in Fig.~\ref{N-MTmin-SD}.b.

The GHAM amplitudes as functions of energy depend strongly on the average velocity and velocity dispersion (see Figs.~\ref{Rate-GHAM-m100}, \ref{Rate-GHAM-m7} and \ref{GHAM-EInt}). Work beyond the scope of this paper is necessary to determine how best aberration features could be used to extract the characteristics of the local WIMP velocity distribution.

\begin{acknowledgments}
G.G. and N.B.  were supported in part by the US Department of Energy Grant DE-FG03-91ER40662, Task C. P.G. was  supported  in part by  the NFS grant PHY-1068111 at the University of Utah. P.G. thanks the Oskar Klein Centre at the University of Stockholm, the Korean Institute for Advanced Studies, and Seoul National University for support during his sabbatical visits.
\end{acknowledgments}


\begin{thebibliography}{99}


\bibitem{Drukier}
A.~K.~Drukier, K.~Freese and D.~N.~Spergel, {\it Detecting cold dark-matter candidates}, Phys.\ Rev.\ D {\bf 33}, 3495 (1986);
K.~Freese, J.~Frieman and A.~Gould, {\it Signal modulation in cold-dark-matter detection}, Phys.\ Rev.\ D {\bf 37}, 3388 (1988).


\bibitem{Ahlen:2009ev}
S.~Ahlen {\it et al.},
{\it The case for a directional dark matter detector and the status of current experimental efforts},
Int.\ J.\ Mod.\ Phys.\ A {\bf 25}, 1-51 (2010)
[arXiv:0911.0323 [astro-ph.CO]].


\bibitem{Spergel}
D.~N.~Spergel, {\it Motion of the Earth and the detection of weakly interacting massive particles},
Phys.\ Rev.\ D {\bf 37}, 1353 (1988).


\bibitem{Copi:1999pw}
  C.~J.~Copi, J.~Heo, L.~M.~Krauss,
  {\it Directional sensitivity, WIMP detection, and the galactic halo},
  Phys.\ Lett.\  B {\bf 461}, 43-48 (1999) [hep-ph/9904499];
  C.~J.~Copi, L.~M.~Krauss,
  {\it Angular signatures for galactic halo WIMP scattering in direct detectors: Prospects and challenges},
  Phys.\ Rev.\  D {\bf 63}, 043507 (2001)
  [astro-ph/0009467];
  B.~Morgan, A.~M.~Green,
  {\it Directional statistics for WIMP direct detection II: 2D read-out},
  Phys.\ Rev.\  D {\bf 72}, 123501 (2005)
  [astro-ph/0508134];
  C.~J.~Copi, L.~M.~Krauss, D.~Simmons-Duffin, S.~R.~Stroiney,
  {\it Assessing alternatives for directional detection of a wimp halo},
  Phys.\ Rev.\  D {\bf 75}, 023514 (2007)
  [astro-ph/0508649];


\bibitem{Alenazi-Gondolo:2008} M.~S.~Alenazi and P.~Gondolo,
{\it Directional recoil rates for WIMP direct detection},
Phys.\ Rev.\ D {\bf 77}, 043532 (2008).


\bibitem{Green&Morgan:2010}
  A.~M.~Green and B.~Morgan,
  {\it The median recoil direction as a WIMP directional detection signal},
  Phys.\ Rev.\  D {\bf 81}, 061301 (2010)
  [arXiv:1002.2717 [astro-ph.CO]].


\bibitem{Green:2010gw}
  A.~M.~Green,
  {\it Dependence of direct detection signals on the WIMP velocity distribution},
  JCAP {\bf 1010}, 034 (2010) [arXiv:1009.0916 [astro-ph.CO]].

\bibitem{Morgan:2005}
B.~Morgan, A.~M.~Green, N.~J.~C.~Spooner,
  {\it Directional statistics for WIMP direct detection},
  Phys.\ Rev.\  D {\bf 71}, 103507 (2005)
  [astro-ph/0408047].


\bibitem{DRIFT}
G.~J.~Alner {\it et al.} [DRIFT Collaboration], {\it The DRIFT-I dark matter detector at Boulby: design, installation and operation}, Nucl.\ Instrum.\ Meth.\ A {\bf 535}, 644 (2004);
B.~Morgan {\it et al.} [DRIFT Collaboration], {\it DRIFT: a directionally sensitive dark matter detector}, Nucl.\ Instrum.\ Meth.\ A {\bf 513}, 226 (2003).


\bibitem{DMTPC}
D.~Dujmic {\it et al.} [DMTPC Collaboration], {\it Observation of the 'head-tail' effect in nuclear recoils of low-energy neutrons}, Nucl.\ Instrum.\ Meth.\ A {\bf 584}, 327 (2008);
A.~Kaboth {\it et al.} [DMTPC Collaboration], {\it A Measurement of Photon Production in Electron Avalanches in CF4}, Nucl.\ Instrum.\ Meth.\ A {\bf 592}, 63 (2008).


\bibitem{NEWAGE}
K.~Miuchi {\it et al.} [NEWAGE Collaboration], {\it Direction-sensitive dark matter search results in a surface laboratory}, Phys.\ Lett.\ B {\bf 654}, 58 (2007);
T.~Tanimori {\it et al.}, {\it Detecting the WIMP-wind via spin-dependent interactions}, Phys.\ Lett.\ B {\bf 578}, 241 (2004).

\bibitem{MIMAC}
D.~Santos {\it et al.}, {\it MIMAC: A Micro-TPC Matrix of Chambers for direct detection of Wimps}, J.\ Phys.\ Conf.\ Ser.\ {\bf 65}, 012012 (2007).


\bibitem{Billard:2009mf}
J.~Billard, F.~Mayet, J.~F.~Macias-Perez and D.~Santos,
{\it Directional detection as a strategy to discover Galactic Dark Matter},
Phys.\ Lett.\ B {\bf 691}, 156 (2010) [arXiv:0911.4086 [astro-ph.CO]].


\bibitem{Ring-like}
N.~Bozorgnia, G.~Gelmini and P.~Gondolo, {\it Ring-like features in directional dark matter detection}, JCAP {\bf 1206}, 037 (2012) [arXiv:1111.6361 [astro-ph.CO]].


\bibitem{Lee:2012pf}
S.~K.~Lee and A.~H.~G.~Peter,
{\it Probing the local velocity distribution of WIMP dark matter with directional detectors},
JCAP {\bf 1204}, 029 (2012) [arXiv:1202.5035v1 [astro-ph.CO]].



\bibitem{Billard:2010jh}
J.~Billard, F.~Mayet and D.~Santos,
{\it A Markov chain Monte Carlo analysis to constrain dark matter properties with directional detection},
Phys.\ Rev.\ D {\bf 83}, 075002 (2011) [arXiv:1012.3960 [astro-ph.CO]].


\bibitem{Gondolo:2002}
 P.~Gondolo,
 {\it Recoil momentum spectrum in directional dark matter detectors},
 Phys.\ Rev.\ D {\bf 66}, 103513 (2002).


 \bibitem{Kuhlen} M.~Kuhlen {\it et al.},
  {\it Dark Matter Direct Detection with Non-Maxwellian Velocity Structure},
JCAP {\bf 1002}, 030 (2010) [arXiv:0912.2358v1 [astro-ph.GA]].

\bibitem{RAVE}
 M.~C.~Smith {\it et al.},
 {\it The RAVE Survey: Constraining the Local Galactic Escape Speed},
 Mon.\ Not.\ Roy.\ Astron.\ Soc.\ {\bf 379}, 755 (2007) [astro-ph/0611671].


\bibitem{Kerr-1986}
F.~J.~Kerr and D.~Lynden-Bell,
{\it Review of galactic constants},
Mon.\ Not.\ Roy.\ Astron.\ Soc.\ {\bf 221}, 1023 (1986).


\bibitem{Reid-2009}
M.~J.~Reid et al.,
{\it Trigonometric Parallaxes of Massive Star Forming Regions: VI. Galactic Structure, Fundamental Parameters and Non-Circular Motions},
Astrophys.\ J.\ {\bf 700}, 137 (2009) [arXiv:0902.3913].

 \bibitem{Bovy-2009}
J.~Bovy, D.~W.~Hogg and H.~Rix,
{\it Galactic masers and the Milky Way circular velocity},
Astrophys.\ J.\ {\bf 704}, 1704 (2009) [arXiv:0907.5423].

\bibitem{McMillan-2010}
P.~J.~McMillan and J.~J.~Binney,
{\it The uncertainty in Galactic parameters},
Mon.\ Not.\ Roy.\ Astron.\ Soc.\ {\bf 402}, 934 (2010) [arXiv:0907.4685].

\bibitem{Schoenrich-2010}
R.~Schoenrich, J.~Binney and W.~Dehnen,
{\it Local Kinematics and the Local Standard of Rest},
Mon.\ Not.\ Roy.\ Astron.\ Soc.\ {\bf 403} 1829 (2010) [arXiv:0912.3693].


\bibitem{Lewin-1996}
J.~D.~Lewin and P.~F.~Smith,
{\it Review of mathematics, numerical factors, and corrections for dark matter experiments based on elastic nuclear recoil},
Astropart.\ Phys.\ {\bf 6}, 87 (1996).

\bibitem{Green-2003} A.~M.~Green,
{\it Effect of realistic astrophysical inputs on the phase and shape of the WIMP annual modulation signal},
Phys.\ Rev.\ D {\bf 68}, 023004 (2003) [arXiv:astro-ph/0304446].

\bibitem{Lang} K.~R.~Lang, {\it Astrophysical Formulae}, Springer-Verlag, New York (1999).



\bibitem{Gelmini-Gondolo:2001} G.~Gelmini and P.~Gondolo,
{\it WIMP Annual Modulation with Opposite Phase in Late-Infall Halo Models},
Phys.\ Rev.\ D {\bf 64}, 023504 (2001) [arXiv:hep-ph/0012315].

\bibitem{ChanIV}
N.~Bozorgnia, G.~Gelmini and P.~Gondolo,
{\it Daily modulation due to channeling in direct dark matter crystalline detectors},
Phys.\ Rev.\ D {\bf 84}, 023516 (2011) [arXiv:1101.2876 [astro-ph.CO]].


\bibitem{Helm:1956} R.~Helm,
{\it Inelastic and elastic scattering of 187-Mev electrons from selected even-even nuclei},
Phys.\ Rev.\ {\bf 104}, 1466 (1956).


\bibitem{CrossSection} T.~Tanaka {\it et al.} [The Super-Kamiokande Collaboration],
{\it An Indirect Search for Weakly Interacting Massive Particles in the Sun Using 3109.6 Days of Upward-going Muons in Super-Kamiokande},
Astrophys.\ J.\ {\bf 742} 78 (2011) [arXiv:1108.3384];
R.~Abbasi {\it et al.} [IceCube Collaboration],
{\it Multi-year search for dark matter annihilations in the Sun with the AMANDA-II and IceCube detectors},
arXiv:1112.1840;
S.~Archambault {\it et al.} [The PICASSO Collaboration],
{\it Constraints on Low-Mass WIMP Interactions on 19F from PICASSO},
arXiv:1202.1240.


\bibitem{Weber:2010} M.~Weber and W.~de Boer,
{\it Determination of the Local Dark Matter Density in our Galaxy},
Astron.\ Astrophys.\ {\bf 509}, 25 (2010) [arXiv:0910.4272].


\bibitem{Salucci:2010}
  P.~Salucci, F.~Nesti, G.~Gentile, C.~F.~Martins,
  {\it The dark matter density at the Sun's location},
  Astron.\ Astrophys.\  {\bf 523}, A83 (2010).
  [arXiv:1003.3101 [astro-ph.GA]].

\bibitem{Pato:2010}
  M.~Pato {\it et al.}
  {\it Systematic uncertainties in the determination of the local dark matter density},
  Phys.\ Rev.\  D {\bf 82}, 023531 (2010).
  [arXiv:1006.1322 [astro-ph.HE]].

\bibitem{Catena:2010}
  R.~Catena, P.~Ullio,
  {\it A novel determination of the local dark matter density},
  JCAP {\bf 1008}, 004 (2010) [arXiv:0907.0018 [astro-ph.CO]].

\bibitem{Evans:2000}
N. W. Evans, C. M. Carollo, and P. T. de Zeeuw,
{\it Triaxial haloes and particle dark matter detection},
Mon. Not. R. Astron. Soc. {\bf 318}, 1131 (2000).

\end{thebibliography}
\end{document}